\documentclass{aa}  

\usepackage{graphicx}
\usepackage{txfonts}
\usepackage{multirow} %para hacer multirow
\usepackage[outdir=./figuras/enpdf/]{epstopdf}
\usepackage{diagbox} %para hacer division en diagonal en celdas tablas
\usepackage{setspace}
\usepackage{float} %para usar H en figuras
\usepackage{blindtext}
\usepackage{gensymb}
\usepackage{multirow}
\usepackage[font=small,labelfont=bf]{caption}

\usepackage{enumitem}

\usepackage{caption}
\usepackage{subcaption}

 \usepackage{tabularx}             
    \usepackage{array}           
\usepackage{ amssymb }

\usepackage{natbib,twoopt}
\usepackage[colorlinks,allcolors=blue]{hyperref} %% to avoid \citeads line fills
\usepackage{verbatim}

\begin{document} 

\newcommand{\doce}{$^{12}$CO }
\newcommand{\docep}{$^{12}$CO}
\newcommand{\trece}{$^{13}$CO }
\newcommand{\trecep}{$^{13}$CO}
\newcommand{\dosuno}{$J=2-1$}
\newcommand{\unocero}{$J=1-0$}
\newcommand{\tresdos}{$J=3-2$}

\newcommand{\seiscincocc}{$J$\,=\,6\,(5)\,$-$\,5\,(4) }
\newcommand{\seiscincoccp}{$J$\,=\,6\,(5)\,$-$\,5\,(4)}
\newcommand{\dosunouu}{$J$\,=\,2\,(1)\,$-$\,2\,(1) }
\newcommand{\dosunouup}{$J$\,=\,2\,(1)\,$-$\,2\,(1)}
\newcommand{\seiscincoud}{$J$\,=\,6\,(1,\,6)\,$-$\,5\,(2,\,3) }
\newcommand{\seiscincoudp}{$J$\,=\,6\,(1,\,6)\,$-$\,5\,(2,\,3)}

\newcommand{\cincocuatro}{$J=5-4$}
\newcommand{\cuatrotres}{$J=4-3$}
\newcommand{\dieznueve}{$J=10-9$}
\newcommand{\ochosiete}{$J=8-7$}
\newcommand{\snso}{$J=68-68$}

\newcommand{\tragua}{$J_{Ka,\,Kc}=6_{1,\,6}-5_{2,\,3}$}
\newcommand{\trsou}{$J_{N}=2_{2}-1_{1}$}
\newcommand{\trsod}{$J=20\,(20)-19\,(20)$}
\newcommand{\trsot}{$J=22\,(22)-21\,(22)$}
\newcommand{\trsoc}{$J_{N}=6_{5}-5_{4}$}
\newcommand{\trsodos}{$J_{Ka,\,Kc}=4_{2,\,2}-4_{1,\,3}$}

\newcommand{\cdsop}{C$^{17}$O}
\newcommand{\cdoop}{C$^{18}$O}
\newcommand{\aguap}{H$_{2}$O}
\newcommand{\vsio}{$^{29}$SiO}
\newcommand{\tsio}{$^{30}$SiO}
\newcommand{\hctn}{HC$_{3}$N}
\newcommand{\htcn}{H$^{13}$CN}
\newcommand{\hcom}{HCO$^{+}$}
\newcommand{\sod}{SO$_{2}$}
\newcommand{\agua}{H$_{2}$O}

\newcommand{\vcero}{$v=0$}
\newcommand{\vuno}{$v=1$}
\newcommand{\vdos}{$v=2$}
\newcommand{\vseis}{$v=6$}

\newcommand{\kms}{\,km\,s$^{-1}$ }
\newcommand{\kmsp}{\,km\,s$^{-1}$}
\newcommand{\ms}{\,M$_{\odot}$ }
\newcommand{\msp}{\,M$_{\odot}$}
\newcommand{\pagb}{post-AGB }
\newcommand{\pagbp}{post-AGB}
\newcommand{\pagbs}{post-AGBs }
\newcommand{\pagbsp}{post-AGBs}

\newcommand{\alls}{AC\,Her, the Red\,Rectangle, 89\,Her, HD\,52961, IRAS\,19157$-$0257, IRAS\,18123+0511, IRAS\,19125+0343, AI\,CMi, IRAS\,20056+1834, and R\,Sct}
\newcommand{\allsp}{AC\,Her, the Red\,Rectangle, 89\,Her, HD\,52961, IRAS\,19157$-$0257, IRAS\,18123+0511, IRAS\,19125+0343, AI\,CMi, IRAS\,20056+1834, and R\,Sct }
\newcommand{\on}{89\,Her }
\newcommand{\onp}{89\,Her}
\newcommand{\iras}{IRAS\,19125+0343 }
\newcommand{\irasp}{IRAS\,19125+0343}
\newcommand{\ac}{AC\,Her }
\newcommand{\acp}{AC\,Her}
\newcommand{\rs}{R\,Sct }
\newcommand{\rsp}{R\,Sct}
\newcommand{\rr}{Red\,Rectangle }
\newcommand{\rrp}{Red\,Rectangle}
\newcommand{\ai}{AI\,CMi }
\newcommand{\aip}{AI\,CMi}

\newcommand{\nir}{NIR excess }
\newcommand{\nirp}{NIR excess}

\newcommand{\fig}{Fig.\,}
\newcommand{\figs}{Figs.\,}
\newcommand{\tab}{Table\,}
\newcommand{\tabs}{Tables\,}

\newcommand{\eq}{Eq.\,}
\newcommand{\eqs}{Eqs.\,}
\newcommand{\sect}{Sect.\,}
\newcommand{\sects}{Sects.\,}
\newcommand{\app}{App.\,}
\newcommand{\secp}{\mbox{\rlap{.}$''$}}

\newcommand{\x}{\,$\times$\,}
\newcommand{\xd}[1]{10$^{#1}$}

\newcommand{\mm}{\,$\pm$\,}

\newcommand{\me}{\textless\,}

\newcommand{\lsim}{\raisebox{-.4ex}{$\stackrel{\sf <}{\scriptstyle\sf \sim}$}}
\newcommand{\gsim}{\raisebox{-.4ex}{$\stackrel{\sf >}{\scriptstyle\sf \sim}$}}

\newcommand\sz{0.90}

   \title{Chemistry of nebulae around binary post-AGB stars: \\ A molecular survey of mm-wave lines \,\thanks{Based on observations with the 30\,m\,IRAM and the 40\,m\,Yebes telescopes. IRAM   is supported by INSU/CNRS (France), MPG (Germany), and IGN (Spain).} \thanks{Final spectra are available at the CDS via anonymous FTP to \url{cdsarc.u-strasbg.fr} (130.79.128.5) or via \url{http://cdsweb.u-strasbg.fr/cgi-bin/gcat?J/A+A/}}}
   \author{I. Gallardo Cava\,\inst{1,2}, V. Bujarrabal\,\inst{2}, J. Alcolea\,\inst{1}, M. Gómez-Garrido\,\inst{2,3}, and M.\,Santander-García\,\inst{1}}     

\institute{Observatorio Astronómico Nacional (OAN-IGN), Alfonso XII 3, 28014, Madrid, Spain\\                 \email{i.gallardocava@oan.es}
        \and
        Observatorio Astronómico Nacional (OAN-IGN), Apartado 112, 28803, Alcalá de Henares, Madrid,     Spain   
        \and
        Centro de Desarrollos Tecnológicos, Observatorio de Yebes (IGN), 19141, Yebes, Guadalajara,              Spain
}

\titlerunning{Chemistry of nebulae around binary post-AGB stars: A molecular survey of mm-wave lines}
\authorrunning{Gallardo Cava, I. et al.}

   \date{}
   \date{Received 30 September 2021 / Accepted 21 January 2022}
% \abstract{}{}{}{}{} 
% 5 {} token are mandatory
 
  \abstract
   {There is a class of binary post-asymptotic giant branch (post-AGB) stars that exhibit remarkable near-infrared (NIR) excess. Such stars  are surrounded by Keplerian or quasi-Keplerian disks, as well as extended outflows composed of gas escaping from the disk. This class can be subdivided into disk- and outflow-dominated sources, depending on whether it is the disk or the outflow that represents most of the nebular mass, respectively. The chemistry of this type of source has been practically unknown thus far.}
   {Our objective is to study the molecular content of nebulae around binary \pagb stars that show disks with Keplerian dynamics, including molecular line intensities, chemistry, and abundances.}
   {We focused our observations on the 1.3, 2, 3\,mm bands of the 30\,m\,IRAM telescope and on the 7 and 13\,mm bands of the 40\,m\,Yebes telescope. Our observations add up $\sim$\,600 hours of telescope time.
   We  investigated the integrated intensities of pairs of molecular transitions for CO, other molecular species, and IRAS fluxes at 12, 25, and 60\,$\mu$m.
   Additionally, we studied isotopic ratios, in particular $^{17}$O\,/\,$^{18}$O, to analyze the initial stellar mass, as well as \docep\,/\,\trecep, to  study the line and abundance ratios.}
   {We present the first single-dish molecular survey of mm-wave lines in nebulae around binary \pagb stars. We conclude that the  molecular content is relatively low in nebulae around binary \pagb stars, as their molecular lines and abundances are especially weaker compared with AGB stars. This fact is very significant in those sources where the Keplerian disk is the dominant component of the nebula. The study of their chemistry allows us to classify nebulae around \acp, the \rrp, \aip, \rsp, and IRAS\,20056+1834 as O-rich, while that of \on is probably C-rich. The calculated abundances of the detected species other than CO are particularly low compared with AGB stars. The initial stellar mass derived from the $^{17}$O\,/\,$^{18}$O ratio for the \rr and \on is compatible with the central total stellar mass derived from previous mm-wave interferometric maps.
The very low \doce\,/\,\trece ratios found in binary \pagb stars reveal a high \trece abundance compared to AGB and other \pagb stars.}
   {}

   %\keywords{Stars: AGB, post$-$AGB, rotating disks, winds, outflows $-$ Stars: Individual: 89\,Her, IRAS\,19125+0343, AC\,Her and R\,Sct}
   
%   giant planet formation --  $\kappa$-mechanism -- stability of gas spheres

   \maketitle
%
%-------------------------------------------------------------------

\section{Introduction}
\label{introduccion}

The spectacular late evolution of low- and intermediate-mass stars (main  sequence  masses  in  the  approximate range of 0.8\,$-$\,8\msp) is characterized by a very copious loss of mass during the asymptotic giant branch (AGB) phase. These kinds of stars experience mass loss rates up to \xd{-4}\msp\,a$^{-1}$ that dominate the evolution in this phase\footnote{We follow the recommendations for units of the IAU Style Manual \citep{wilkins1990,wilkins1995}.  Therefore, we use the term annus, abbreviated as ``a'', for year.}. 
The ejected material creates an expanding circumstellar envelope (CSE).
These stars are considered one of the most relevant contributors of dust and enriched material to the Interstellar Medium (ISM); see \citet{gehrz1989, matsuura2009}.
This environment is favorable for the formation of simple molecules and dust.
We find three different kinds of envelopes depending on the C/O ratio: the M-type stars with C/O\,<\,1, the S-type stars with C/O\,$\approx$\,1, and C-type with C/O\,>\,1.
The kind of molecules and dust grains found in CSEs is determined by the C/O ratio. We find O-bearing molecules, such as SiO or H$_{2}$O, and silicate dust in M-type stars, making the envelope of these stars O-rich \citep{engels1979, velillaprieto2017}. On the contrary, we find C-bearing molecules, such as HCN and CS, often alongside SiS, in C-type stars \citep{olofsson1993, cernicharo2000}. Amorphous carbon dust formed out of primary carbon is dominant in C-rich envelopes, while silicon carbide and magnesium sulphide are minor widespread dust components \citep{zhukovskagail2008, massalkhi2020}.

Molecular lines were analysed in a large sample of evolved stars \citep{bujarrabal1994a,bujarrabal1994b}. The authors investigate line ratios between CO and other O- and C-bearing species to analyze the molecular intensities of these sources. They also study O- and C-bearing molecule ratios to find the necessary criteria to discern between O- and C-rich envelopes around evolved stars. They found, in particular, that SiO and HCN lines can be as intense as CO lines in O- and C-rich stars, respectively.

When the star attains the post-asymptotic giant branch (post-AGB) phase, the expanding envelope becomes a pre-planetary nebula (pPN). The final stage of the evolution of the (low- or intermediate-mass) star is heralded when the exposed core, a white dwarf, ionizes the circumstellar nebula and becomes a planetary nebula (PN).
The chemical behaviour in pPNe is more complex than for normal circumstellar envelopes around AGB stars; since O- and C-rich pPNe could present differences in the line ratios compared with O- and C-rich stars \citep{bujarrabal1994a}. For example, SiO is systematically weaker in pPNe than in evolved stars, while HCN appears to show the opposite trend.

The chemical evolution during the AGB to pPN to PN phases has been investigated in several surveys \citep{bujarrabal1994a,bujarrabal1994b,bujarrabal2006,cernicharo2011}. The star experiences striking changes in the molecular composition at the pPN phase \citep{pardo2007, park2008, zhang2013}.
However, the chemistry in pPN is poorly studied, even molecular line surveys of PNe are relatively scarce  \citep{zhang2017}.

There is a kind of binary post-AGB stars (binary systems including a \pagb star) that systematically shows evidence for the presence of disks orbiting the central stars \citep{vanwinckel2003,ruyter2006,bujarrabal2013a,hillen2017}. All of them present remarkable near-infrared (NIR) excess and the narrow CO line profiles characteristic of rotating disks.
The presence of hot dust close to the stellar system is suspected from their spectral energy distribution (SED) and its disk-like shape has been confirmed by interferometric IR data \citep{hillen2017,kluska2019}.
The IR spectra of these sources reveal the presence of highly processed grains, which implies that their disks must be stable structures \citep{jura2003,sahai2011,gielen2011a}.
The low-$J$ rotational lines of CO have been well studied in sources with NIR excess \citep{bujarrabal2013a}. 
They show narrow CO lines and relatively wide wings. These line profiles are similar to those in young stars surrounded by a rotating disk made of remnants of interstellar medium and those expected from disk-emission \citep{bujarrabal2005,guilloteau2013}. These results indicate that the CO emission lines of our sources come from Keplerian (or quasi-Keplerian) disks.

Due to the relatively small size of these disks, deep  studies of Keplerian disks around \pagb stars need high angular- and spectral-resolution. To date, there have been only four well-resolved Keplerian disks identified: in the Red\,Rectangle \citep{bujarrabal2013b,bujarrabal2016}, AC\,Her \citep{bujarrabal2015, gallardocava2021}, IW\,Car \citep{bujarrabal2017}, and IRAS\,08544$-$4431 \citep{bujarrabal2018}. 
These sources are disk-dominated, because the disks contain most of the material of the nebula ($\sim$\,90\% of the total mass),
while the rest of the mass ($\sim$10\%), corresponds to an extended bipolar outflow surrounding the disk.
Very detailed studies \citep{gallardocava2021} also show strong indications of the presence of Keplerian disks in \onp, \irasp, and \rsp.
The nebulae around \iras and \rs are dominated by the expanding component, instead of the Keplerian disk, because $\sim$\,75\% of the total mass is located in the outflow. These sources belong to the outflow-dominated subclass. \on is in an intermediate case in between the disk- and the outflow-dominated sources, because $\sim$\,50\% of the material of the nebula is in the Keplerian disk.

There are other binary \pagb stars that are surrounded by a Keplerian disk that have only been studied with low angular resolution, all of them presenting NIR excess and narrow CO line profiles characteristic of rotating disks \citep{bujarrabal2013a}.

The chemistry of this class of binary post-AGB sources with Keplerian disks has been practically unknown, with only CO having been very well observed. Furthermore, H$^{13}$CN, \textsc{C\,i}, \textsc{C\,ii} were also detected in the \rr and two maser emission lines were found in \aip. In this work, we present a deep and wide survey of radio lines in ten of these sources. All of them have been observed in the 7 and 13\,mm bands, and most of them have also been  observed at 1.3, 2, and 3\,mm.
The scarce literature about the molecular composition in very evolved objects shows that pPNe display dramatic changes in the chemical composition in the molecular gas with respect to that of their AGB circumstellar progenitors. 
Nevertheless, this kind of nebulae surrounding binary \pagb stars seem to present a relatively low molecular content. Here we carry out an analysis to explore whether this lower abundance depends on the specific subclass of the binary \pagb star (disk- or outflow-dominated sources).
Additionally, we want to investigate the possibility of classifying some of our objects as O-\,/\,C-rich, based on line ratios of different molecules.

This paper is laid out as follows. We present our binary \pagb star sample in \sect\ref{fuentes}, where relevant results from previous observations are given. Technical information on our observations is presented in \sect\ref{obs} for both telescopes (30\,m\,IRAM and 40\,m\,Yebes). We provide our spectra in \sect\ref{resultados}. Detailed discussions about molecular intensities, chemistry, abundances, and isotopic ratio analysis can be found in \sect\ref{discusiones}. Finally, we summarize our conclusions in \sect\ref{conclusiones}.

\begin{figure}[H]
\centering
%width=\sz\linewidth
\includegraphics[width=\sz\linewidth]{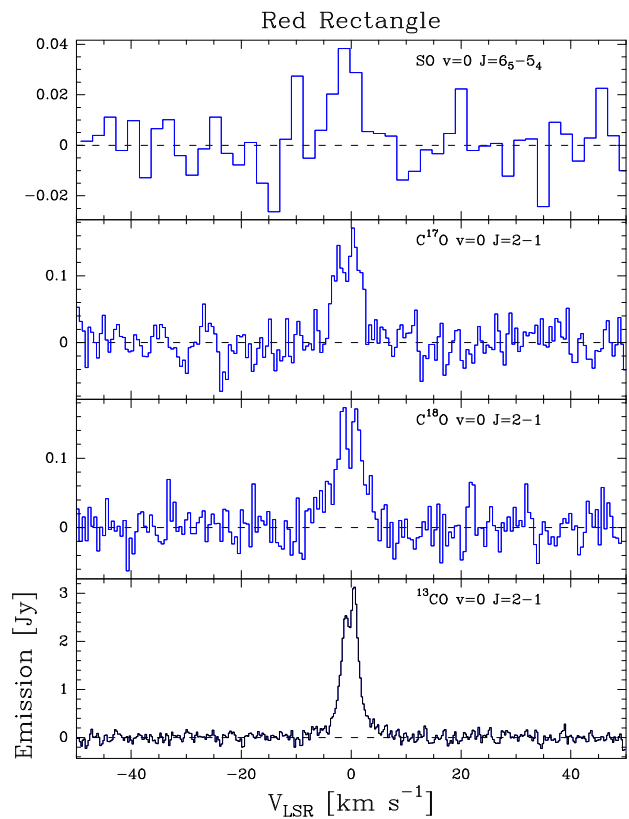}
\caption{\small Spectra of the newly detected transitions in the \rrp. For comparison purposes, we also show the \trece\ \dosuno\ line in black \citep{bujarrabal2013a}. The x-axis indicates velocity with respect to the local standard of rest ($V_{LSR}$) and the y-axis represents the detected flux measured in Jansky.}
    \label{fig:rr_mol}  
\end{figure}

\begin{figure}[H]
\centering
%width=\sz\linewidth
\includegraphics[width=\sz\linewidth]{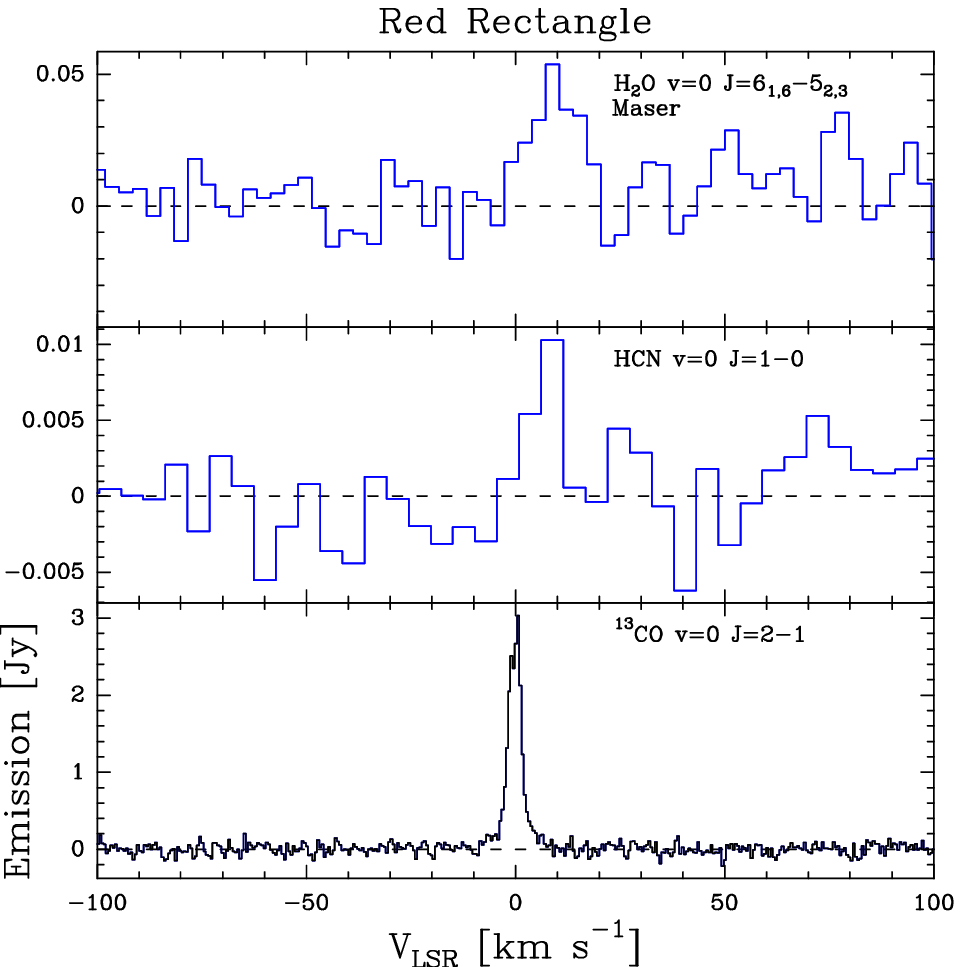}
\caption{\small Spectra of the newly tentative detections in the \rrp. For comparison purposes, we also show the \trece\ \dosuno\ line in black \citep{bujarrabal2013a}. The x-axis indicates velocity with respect to the local standard of rest ($V_{LSR}$) and the y-axis represents the detected flux measured in Jansky.}
    \label{fig:rr_mol_ten}  
\end{figure}

\begin{figure}[H]
\center
\includegraphics[width=\sz\linewidth]{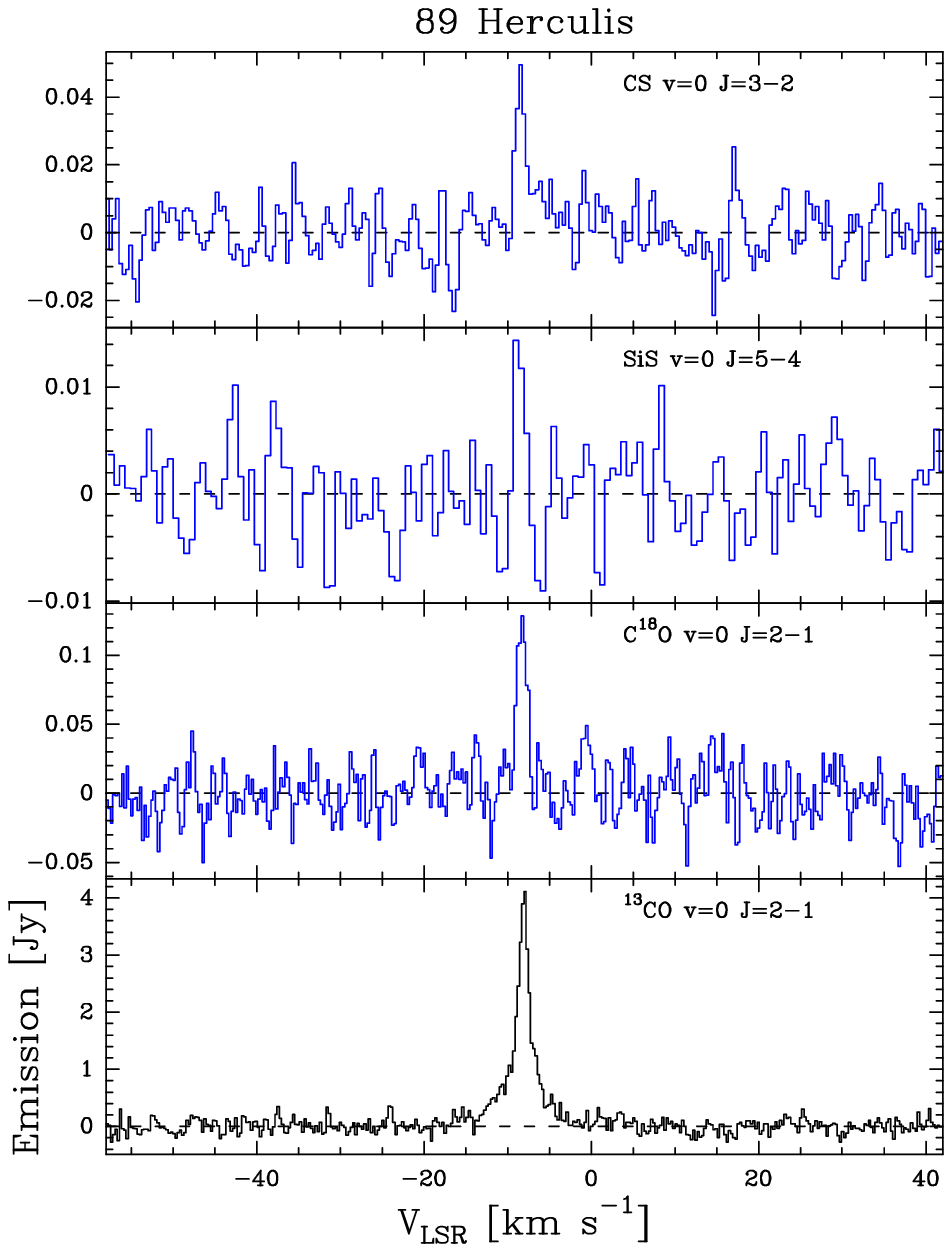}
\caption{\small Spectra of the newly detected transitions in \onp. For comparison purposes, we also show the \trece\ \dosuno\ line in black \citep{bujarrabal2013a}. The x-axis indicates velocity with respect to the local standard of rest ($V_{LSR}$) and the y-axis represents the detected flux measured in Jansky.}
    \label{fig:89her_mol}  
\end{figure}

\begin{figure}[H]
\center
\includegraphics[width=\sz\linewidth]{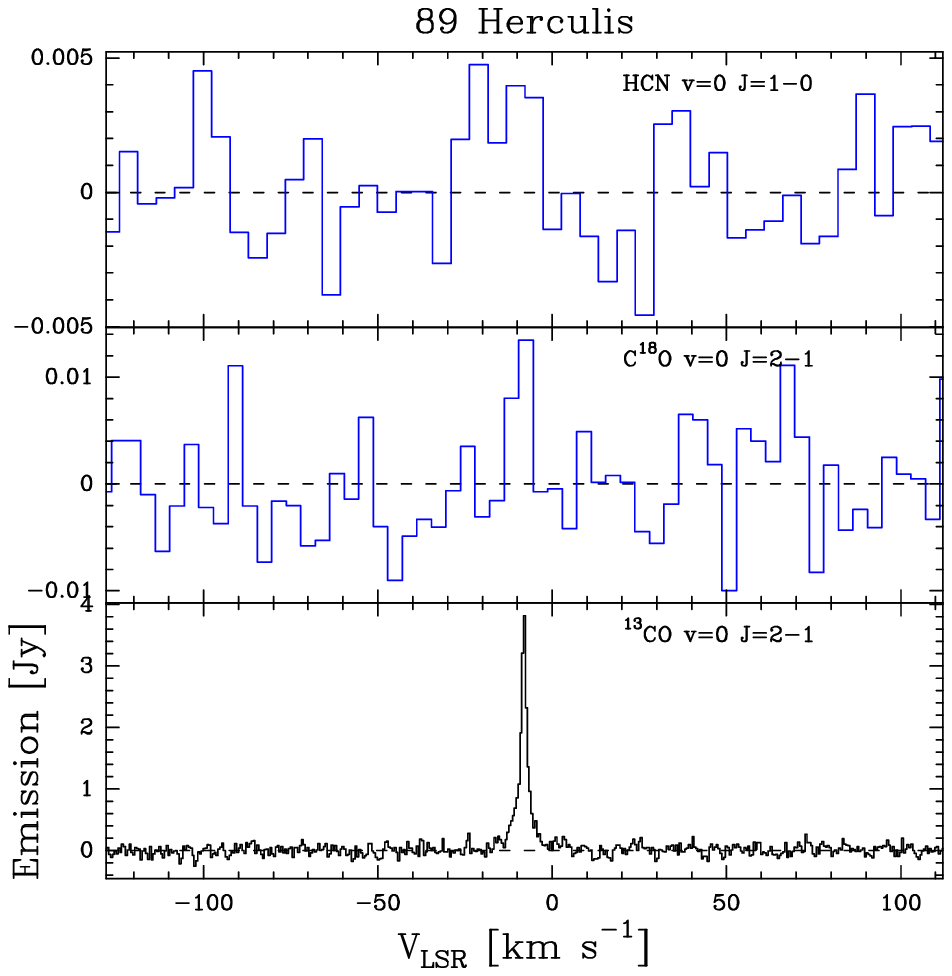}
\caption{\small Spectra of the newly tentative detections in \onp. For comparison purposes, we also show the \trece\ \dosuno\ line in black \citep{bujarrabal2013a}. The x-axis indicates velocity with respect to the local standard of rest ($V_{LSR}$) and the y-axis represents the detected flux measured in Jansky.}
    \label{fig:89her_mol_ten}  
\end{figure}

\begin{figure}[H]
\center
\includegraphics[width=\sz\linewidth]{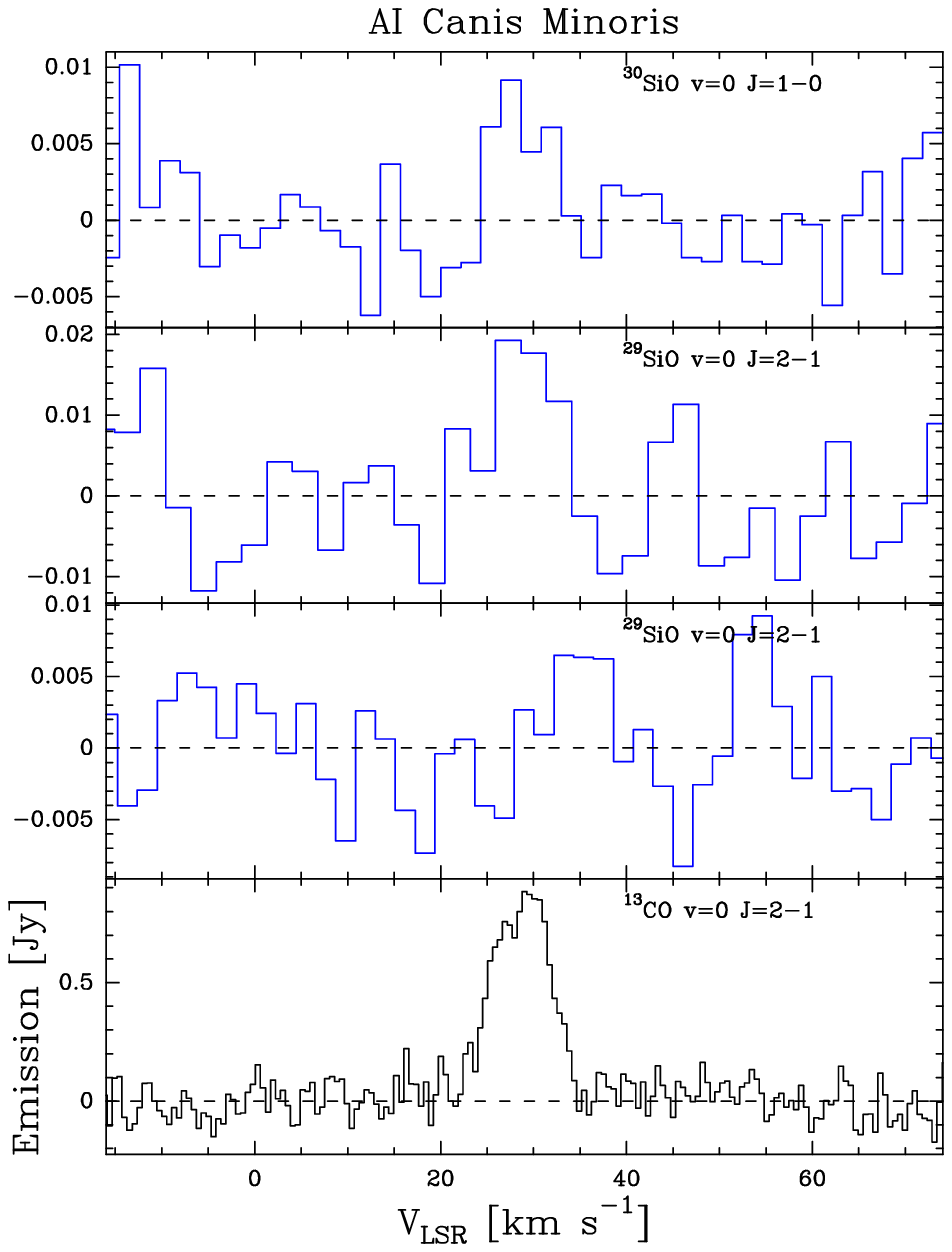}
\caption{\small Spectra of the newly tentative detections in AI\,CMi. For comparison purposes, we also show the \trece\ \dosuno\ line in black \citep{bujarrabal2013a}. The x-axis indicates velocity with respect to the local standard of rest ($V_{LSR}$) and the y-axis represents the detected flux measured in Jansky.}
    \label{fig:aicmi_mol_1_ten}  
\end{figure}
\begin{figure}[H]
\center
\includegraphics[width=\sz\linewidth]{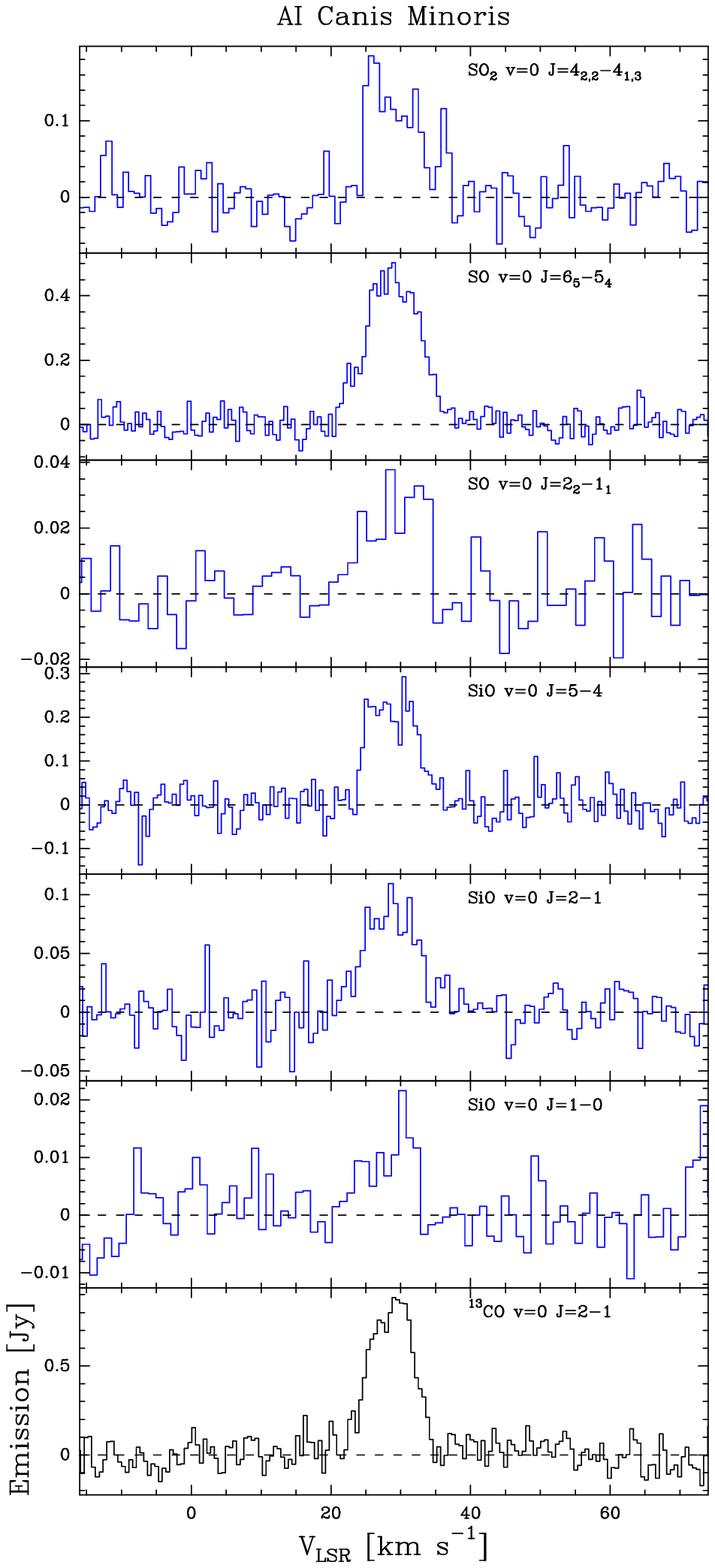}
\caption{\small Spectra of the newly detected transitions in AI\,CMi. For comparison purposes, we also show the \trece\ \dosuno\ line in black \citep{bujarrabal2013a}. The x-axis indicates velocity with respect to the local standard of rest ($V_{LSR}$) and the y-axis represents the detected flux measured in Jansky.}
    \label{fig:aicmi_mol_1}  
\end{figure}

\begin{figure}[H]
\center
\includegraphics[width=\sz\linewidth]{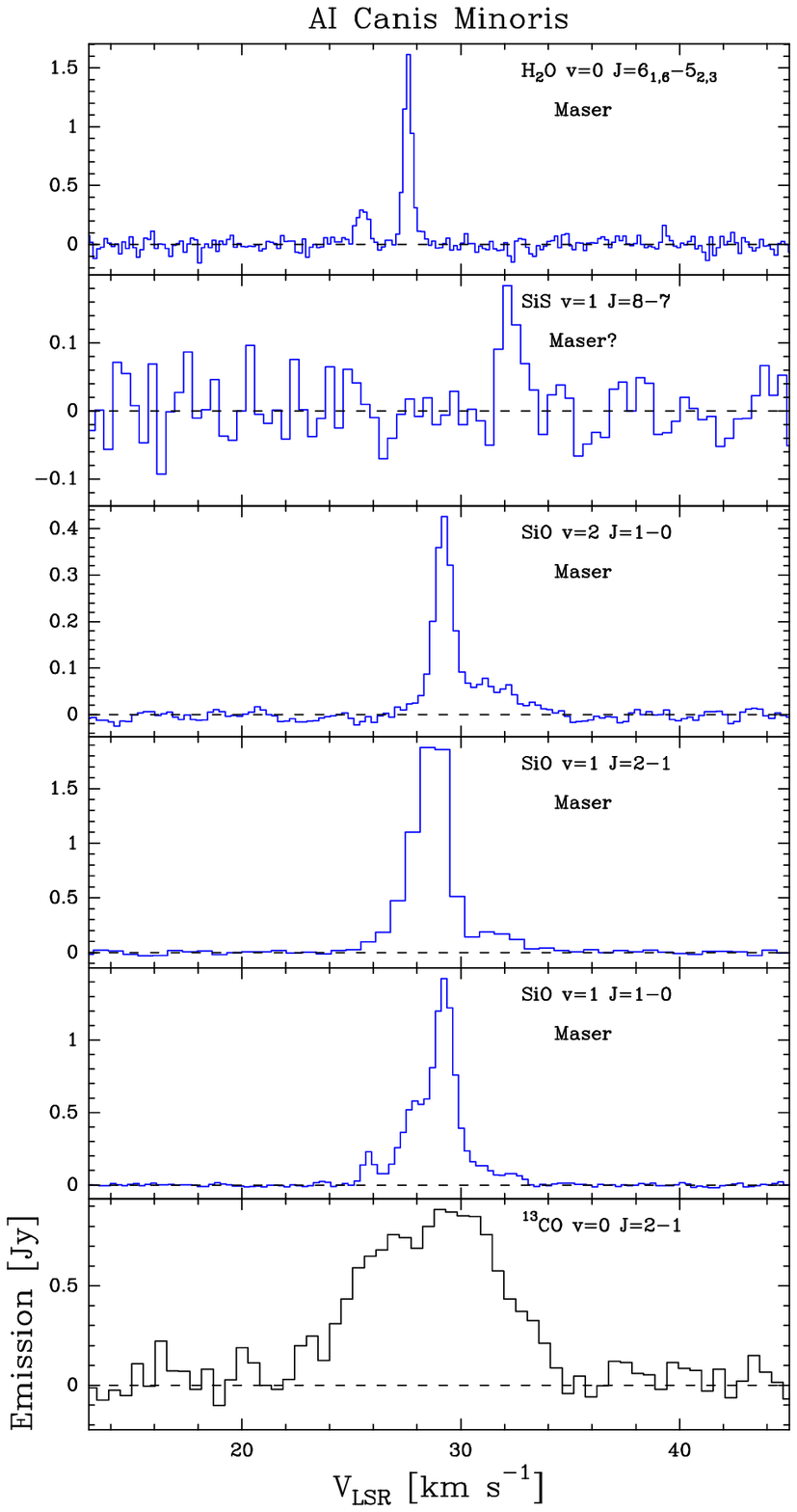}
\caption{\small Spectra of the newly maser detections in AI\,CMi. For comparison purposes, we also show the \trece\ \dosuno\ line in black \citep{bujarrabal2013a}. The x-axis indicates velocity with respect to the local standard of rest ($V_{LSR}$) and the y-axis represents the detected flux measured in Jansky.}
    \label{fig:aicmi_mol_2}  
\end{figure}

\begin{figure}[H]
\center
\includegraphics[width=\sz\linewidth]{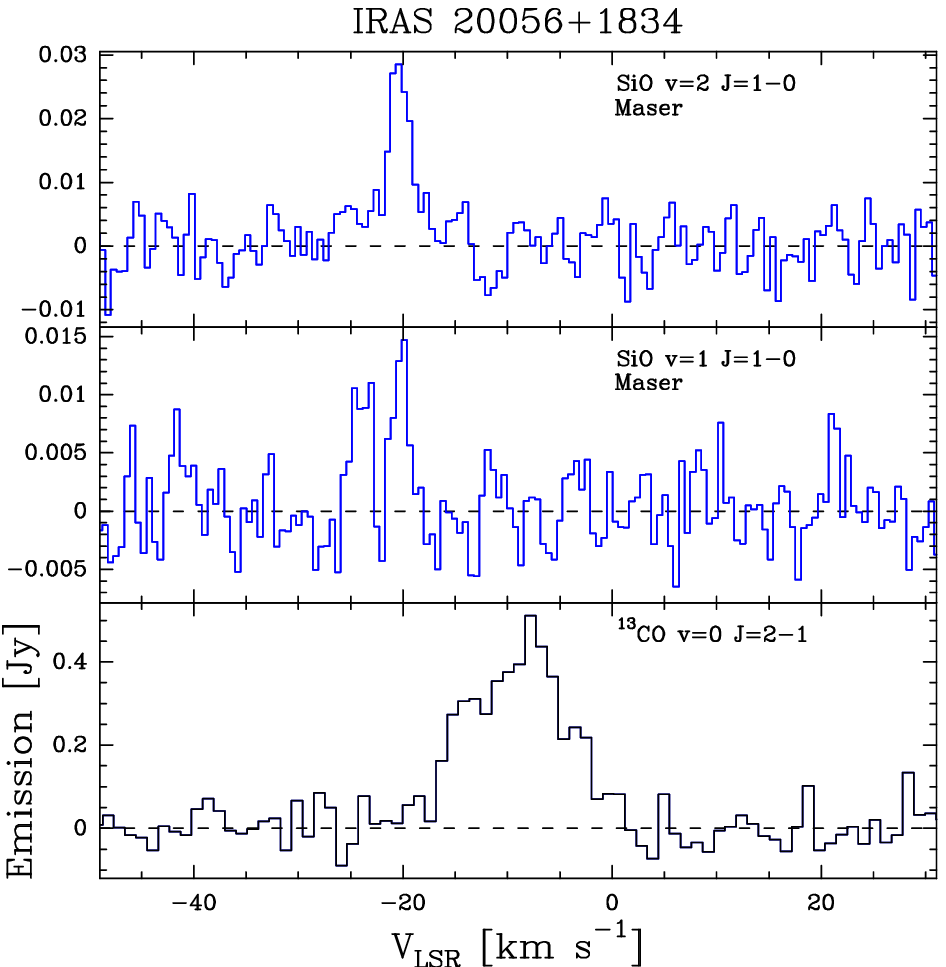}
\caption{\small Spectra of the newly maser detections in IRAS\,20056+1834. For comparison purposes, we also show the \trece\ \dosuno\ line in black \citep{bujarrabal2013a}. The x-axis indicates velocity with respect to the local standard of rest ($V_{LSR}$) and the y-axis represents the detected flux measured in Jansky.}
    \label{fig:iras20056_mol}  
\end{figure}

\begin{figure}[H]
\center
\includegraphics[width=\sz\linewidth]{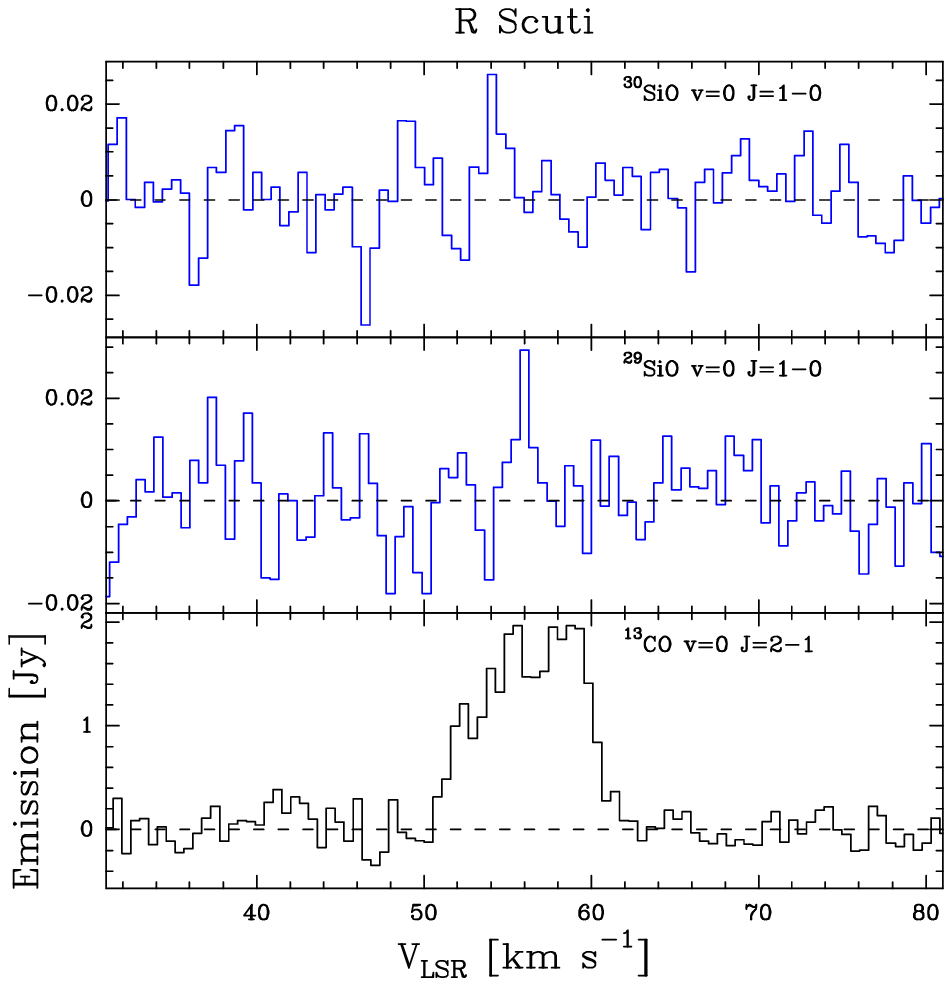}
\caption{\small Spectra of the newly tentative detections in \rsp. For comparison purposes, we also show the \trece\ \dosuno\ line in black \citep{bujarrabal2013a}. The x-axis indicates velocity with respect to the local standard of rest ($V_{LSR}$) and the y-axis represents the detected flux measured in Jansky.}
    \label{fig:rsct_mol_1_ten}  
\end{figure}
\begin{figure}[H]
\center
\includegraphics[width=\sz\linewidth]{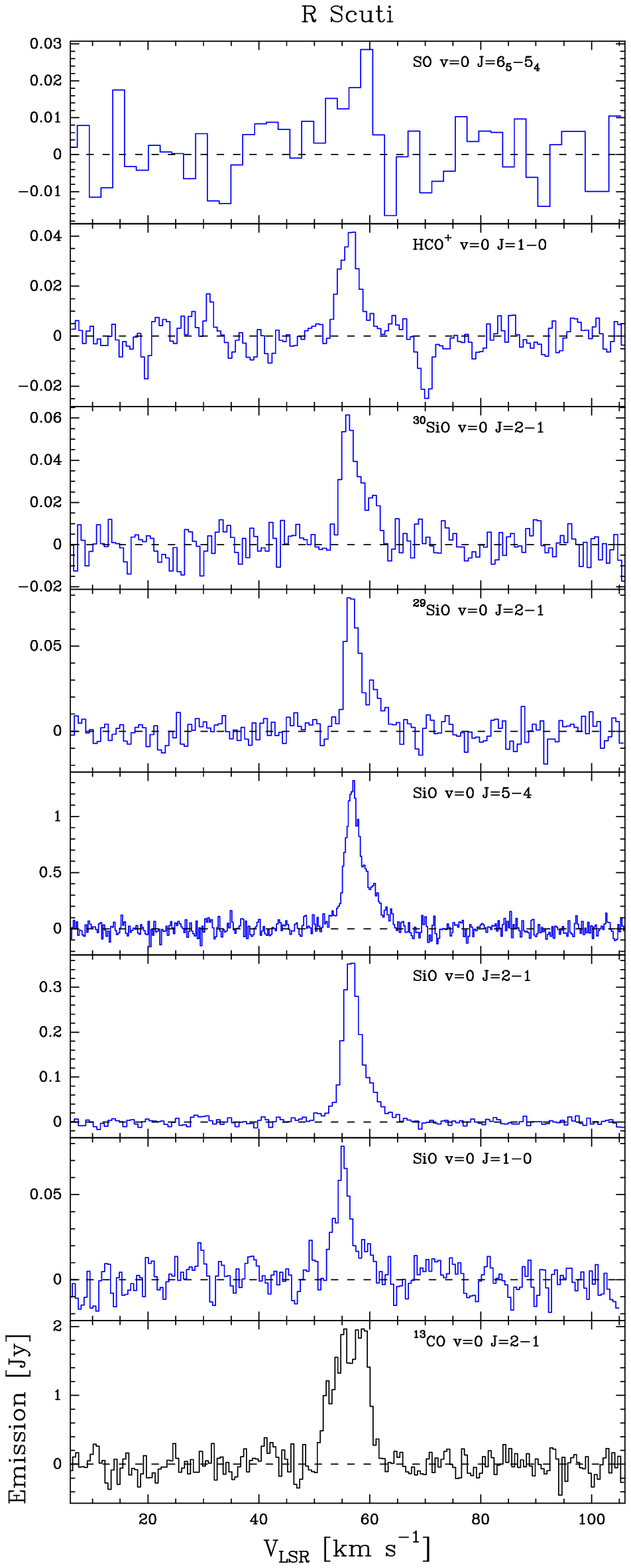}
\caption{\small Spectra of the newly detected transitions in \rsp. For comparison purposes, we also show the \trece\ \dosuno\ line in black \citep{bujarrabal2013a}. The x-axis indicates velocity with respect to the local standard of rest ($V_{LSR}$) and the y-axis represents the detected flux measured in Jansky.}
    \label{fig:rsct_mol_1}  
\end{figure}

\begin{figure}[H]
\center
\includegraphics[width=\sz\linewidth]{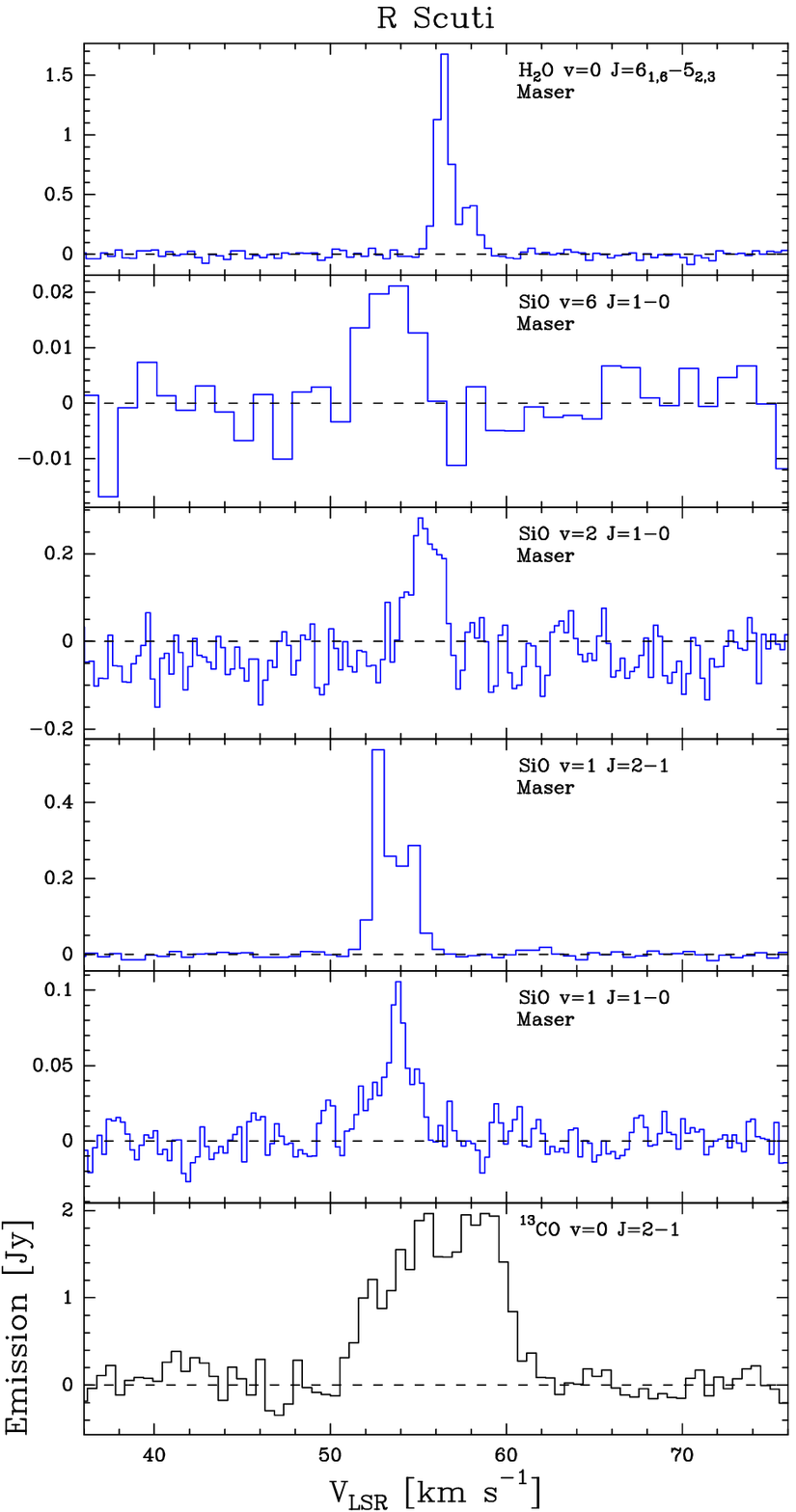}
\caption{\small Spectra of the newly maser detections in \rsp. For comparison purposes, we also show the \trece\ \dosuno\ line in black \citep{bujarrabal2013a}. The x-axis indicates velocity with respect to the local standard of rest ($V_{LSR}$) and the y-axis represents the detected flux measured in Jansky.}
    \label{fig:rsct_mol_2}  
\end{figure}

\begin{figure}[h]
\center
\includegraphics[width=\sz\linewidth]{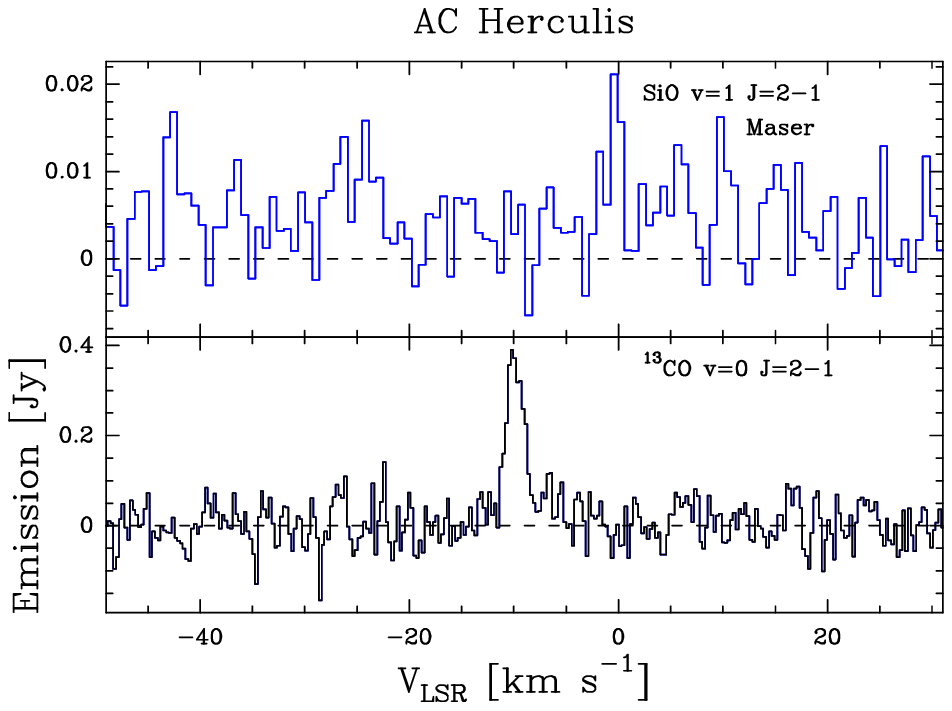}
\caption{\small Spectra of the newly tentative maser detection in \acp. For comparison purposes, we also show the \trece\ \dosuno\ line in black \citep{bujarrabal2013a}. The x-axis indicates velocity with respect to the local standard of rest ($V_{LSR}$) and the y-axis represents the detected flux measured in Jansky.}
    \label{fig:acher_mol_1}  
\end{figure}

\begin{table}[h]
\caption{Binary \pagb stars observed in this paper.}
\small
%\tiny

%\centering
\vspace{-5mm}
\begin{center}
\begin{tabular}{lcccc}
\hline \hline
\noalign{\smallskip}
%&&&&&\multicolumn{4}{|c|}{}\\
 %\\[-2ex]

\multirow{2}{*}{Source} & $M_{neb}$   & $\frac{Disk}{Total}$  & $V_{LSR}$ & $d$  \\
  & [M$_{\odot}$] & [\%]  & [km\,s$^{-1}$] & [pc]   \\
\hline
\\[-2ex]
\vspace{1mm} 

AC\,Herculis  &  8.3\x\xd{-4}  & \gsim\,95  & $-$9.7 & 1100  \\
\vspace{1mm}
Red\,Rectangle & 1.4\x\xd{-2} & 90 & 0 & 710 \\
\vspace{1mm} 
89\,Herculis   & 1.4\x\xd{-2} & 50 & $-$8.0 & 1000  \\
\vspace{1mm}
HD\,52961  & 1.3\x\xd{-2} & $\sim$\,50 & $-$7 & 2800 \\
\vspace{1mm} 
IRAS\,19157$-$0257 & 1.4\x\xd{-2} & $\sim$\,50 & 46 & 2900  \\
\vspace{1mm} 
IRAS\,18123+0511 & 4.7\x\xd{-2} & $\sim$\,30 & 99 & 3500  \\
\vspace{1mm} 
IRAS\,19125+0343 & 1.1\x\xd{-2} & 30 & 82.0 & 1500   \\ 
\vspace{1mm}
AI\,Canis\,Minoris & 1.9\x\xd{-2} & 25 & 29 & 1500  \\
\vspace{1mm}
IRAS\,20056+1834 & 1.0\x\xd{-1} & 25 & $-$9 & 3000  \\
%\vspace{1mm} 
R\,Scuti & 3.2\x\xd{-2}  & 25 & 56.1  & 1000  \\
\hline
\end{tabular}
% \\[1ex]

\end{center}
\small
\vspace{-1mm}
\textbf{Notes.} 
Sources are ordered based on their outflow\,/\,disk mass ratio.
Nebular masses ($M_{neb}$) and velocities ($V_{LSR}$) are derived from our single-dish and interferometric observations \citep{bujarrabal2013a,bujarrabal2016, gallardocava2021}, except for HD\,52961 (Gallardo\,Cava et al. in prep).
Distances ($d$) are adopted from \citet{bujarrabal2013a} except for HD\,52961 \citep{oomen2019}.

\label{prop}
\end{table}

\section{Description of the sources and previous results}
\label{fuentes}

We performed single-dish observations of ten sources (\tab\ref{prop}) using the 30\,m\,IRAM and the 40\,m\,Yebes telescopes. These sources are identified as binary \pagb stars with far-infrared (FIR) excess that is indicative of material ejected by the star. All of them also show significant NIR excess (see \fig\ref{fig:diagramacolorcolor}) characteristic of rotating disks \citep[][]{oomen2018}. They have been poorly studied in the overall search for molecules other than CO.
We adopted the distances used in \citet{bujarrabal2013a}. We refrained from measuring distances via parallax measurements, because in the case of binary stars, it is overly complex \citep{dominik2003}.
The velocities are derived from the single-dish and interferometric observations \citep{bujarrabal2013a,bujarrabal2015,bujarrabal2016,bujarrabal2017,bujarrabal2018, gallardocava2021}.

\begin{table}[h]
\caption{Observed frequency ranges}
\normalsize
\small
%\tiny
%\LARGE
%\centering
\vspace{-5mm}
\begin{center}
\resizebox{\linewidth}{!}{
\begin{tabular}{lccccc}
\hline \hline
\noalign{\smallskip}
Source &  $\lambda=1.3$\,mm & $\lambda=2$\,mm  & $\lambda=3$\,mm & $\lambda=7$\,mm & $\lambda=13$\,mm \\

\hline
\\[-2ex]
\vspace{1mm} 
AC\,Herculis  &  \checkmark &  \checkmark &  \checkmark &  \checkmark & \checkmark   \\
\vspace{1mm} 
Red\,Rectangle & \checkmark &  \checkmark &  \checkmark &  \checkmark & \checkmark \\
\vspace{1mm} 
89\,Herculis & \checkmark  & \checkmark & \checkmark & \checkmark & \checkmark \\
\vspace{1mm} 
HD\,52961  & \checkmark  & \checkmark & \checkmark & \checkmark & \checkmark \\
\vspace{1mm}
IRAS\,19157$-$0257  & &  \checkmark &  \checkmark &  \checkmark & \checkmark   \\
\vspace{1mm}
IRAS\,18123+0511 &  \checkmark &  \checkmark &  \checkmark & \checkmark & \checkmark \\
\vspace{1mm}
\irasp   & \checkmark  & &  \checkmark & \checkmark & \checkmark \\ 
\vspace{1mm}
AI\,Canis\,Minoris   & \checkmark  & \checkmark  & \checkmark &  \checkmark & \checkmark \\
\vspace{1mm}
IRAS\,20056+1834 &  \checkmark  & & \checkmark & \checkmark & \checkmark  \\
%\vspace{1mm}
R\,Scuti & \checkmark  & \checkmark  & \checkmark & \checkmark & \checkmark \\

\hline
\end{tabular}}
% \\[1ex]
\end{center}
\small
\vspace{-1mm}

\label{rangos}
\end{table}

\begin{table}[h]
\caption{Molecular transitions detected in this work.}
\small
%\tiny
%\centering
\vspace{-5mm}
\begin{center}
%\resizebox{\columnwidth}{!}{
\begin{tabular}{lllcl}
\hline \hline
\noalign{\smallskip}
%&&&&&\multicolumn{4}{|c|}{}\\
 %\\[-2ex]

\multirow{2}{*}{Molecule} & \multicolumn{2}{c}{Transition}  & $\nu$ \\
 & Vibrational & Rotational &  [MHz]  \\
\hline
\\[-2ex]

 \vspace{1mm}
C$^{17}$O & \vcero & \dosuno & 224713.53  \\
 \vspace{1mm}
 
C$^{18}$O & \vcero & \dosuno & 219560.36  \\

$^{28}$SiO & \vcero & \unocero & 43423.85  \\
 &  & \dosuno & 86846.99  \\
 &  & \cincocuatro & 217104.98 \\
 &  \vuno & \unocero & 43122.08 \\
 &  & \dosuno & 86243.37   \\ 
 &  \vdos & \unocero & 42820.59  \\ 
  \vspace{1mm} 
 & \vseis & \unocero & 41617.40  \\ 
 
$^{29}$SiO & \vcero & \unocero & 42879.95 \\
 \vspace{1mm} 
 &  & \dosuno & 85759.19 \\

$^{30}$SiO & \vcero & \unocero & 42373.34  \\
 &  & \dosuno & 84746.17 \\
 \vspace{1mm} 
 & \vuno &  \unocero & 42082.47 \\
 
\vspace{1mm}
HCN & \vcero & \unocero & 88630.42 \\
\vspace{1mm} 

HCO$^{+}$ & \vcero & \unocero & 89188.52  \\
\vspace{1mm} 

CS & \vcero & \tresdos & 146969.00  \\

%SiS & \vcero & \dosuno & 36309.63  \\
 SiS & \vcero & \cincocuatro & 90771.56  \\
\vspace{1mm} 
 & \vuno & \ochosiete & 144520.36  \\

SO & \vcero & \trsou & 86093.95  \\
\vspace{1mm} 
 & & \trsoc & 219949.44  \\
 % & & $J=20\,(20)-19\,(20)$ & 144243.27 & IRAM & 17.1 & 200 & 6.7 \\
  
  %& & $J=22\,(22)-21\,(22)$ & 144915.36 & IRAM & 16.9 & 200 & 6.7 \\
 
%SO$_{2}$ & \vcero & \snso & 146605.52 & IRAM & 16.8 & 200 & 6.8 \\
\vspace{1mm} 
 SO$_{2}$ & \vcero & \trsodos & 146605.52 \\

H$_{2}$O & \vcero & \tragua & 22235.08 \\

\hline
\end{tabular}
%}
% \\[1ex]

\end{center}
\small
\vspace{-1mm}
\textbf{Notes.} Rest frequencies are taken from Cologne Database for Molecular Spectroscopy
(CDMS) and Jet Propulsion Laboratory (JPL).

\label{lineas}
\end{table}

\subsection{\acp}
\ac is a binary \pagb star \citep{oomen2019}. Recent mm-wave interferometric observations confirm that \ac presents a Keplerian disk that clearly dominates the nebula. Observational data and models are compatible with a very diffuse outflowing component surrounding the rotating component. The nebula presents a total mass of 8.3\x\xd{-3}\ms and we find that the mass of the outflow must be \lsim\,5\%. The rotation of the disk is compatible with a central total stellar mass of $\sim$\,1\ms \citep{gallardocava2021}.
The molecular content of this source was unknown, except for the well-studied CO lines \citep{bujarrabal2013a, gallardocava2021}.

\subsection{\rrp}

The \rr is the best-studied object of our sample. It is a binary \pagb star \citep{oomen2019}, 
with an accretion disk that emits in the UV \citep{witt2009, thomas2013}. This UV emission excite a central \textsc{H\,ii} region \citep{jura1997,bujarrabal2016} and must also yield a Photo Dissociation Region (or Photon-Dominated Regions, or PDR) between the \textsc{H\,ii} region and the extended Keplerian disk.
%45555555555555555555555555555555555555

Its mm-wave interferometric maps reveal that the nebula contains a rotating disk, containing 90\% of the total nebular mass  \citep[1.4\x\xd{-2}\msp, see][]{bujarrabal2016}, together with an expanding low-mass component.
The CO line profiles of the \rr are narrow and present weak wings \citep{bujarrabal2013a}.
\citet{bujarrabal2016} find high excitation lines: \cdsop\ $J=6-5$ and \htcn\ $J=4-3$. The \cdsop\ line is useful to study regions closer than 60\,AU from the central binary star. The presence of the \htcn\ line means that this molecule is significantly abundant in the central region of the disk, more precisely, closer than 60\,AU. 
The detection of \textsc{C\,i}, \textsc{C\,ii}, H$^{13}$CN, and FIR lines is consistent with the presence of a PDR \citep[see e.g.][]{agundez2008}, because these lines are the best tracers of these regions.

\subsection{\onp}
This source is a binary \pagb star with NIR excess that implies the presence of hot dust \citep{ruyter2006}. The dust is in a stable structure where large dust grains form and settle to the midplane  \citep{hillen2013, hillen2014}.
\on was studied in detail in mm-wave interferometric maps by \citet{gallardocava2021}. The nebula  contains an extended hourglass-like component and a rotating disk in its innermost region.
The total mass of the nebula is 1.4\x\xd{-2}\ms and the outflow mass represents, at least, 50\%. 

The molecular content of this source was very poorly studied, except for the single-dish CO observations, which show narrow lines similar to those of the \rrp, but with more prominent wings \citep{bujarrabal2013a}. 
\subsection{HD\,52961}

This source is a binary \pagb star \citep{gielen2011b,oomen2018}. It presents relatively narrow CO line profiles. Its molecular content was poorly known.
\citet{gielen2011b} found CO$_{2}$ and the fullerene C$_{60}$. According to these authors, this source could be O-rich based on its strong similarities to an other post-AGB disk source (EP\,Lyr).

\subsection{IRAS\,19157$-$0257 and IRAS\,18123+0511}

IRAS\,19157$-$0257 and IRAS\,18123+0511 are also binary \pagb stars \citep{oomen2019,scicluna2020}.

The variability of IRAS\,19157$-$0257 is cataloged as irregular \citep{kiss2007}.
Its CO line profiles show relatively narrow lines that are similar to those of \onp. The nebular mass is 1.3\x\xd{-2}\ms \citep{bujarrabal2013a}. Its molecular content, apart from CO, was unknown.

IRAS\,18123+0511 show wide CO line profiles (similar to those found in \irasp). Its nebular mass is 4.7\x\xd{-2}\ms \citep{bujarrabal2013a}.
No molecular species apart from CO have been detected in these sources despite numerous previous observations \citep{gomez1990,lewis1997,deguchi2012,liujiang2017}. Our new data improve the rms values of other works.

\subsection{\irasp}

\iras is a binary \pagb star \citep{gielen2008}, which also belongs to this class of binary \pagb star with remarkable NIR excess \citep{oomen2019}.
Recent mm-wave interferometric observations and models reveal that the nebula around this binary \pagb star is composed of a rotating disk with Keplerian dynamics and an extended outflowing component around it \citep{gallardocava2021}. The total mass of the nebula is 1.1\x\xd{-2}\ms and the outflow mass represents 70\%.
The CO lines of this source are narrow but present prominent wings \citep{bujarrabal2013a}. Apart from CO, the molecular content of this source was unknown.

\subsection{\aip} \label{aicmi_res_previos}

\ai is an irregular pulsating star with variable amplitude and multiperiodicity that is in an early transition phase from the AGB to the post-AGB stage.
The spectral type at maxima is G5\,$-$\,G8\,I \citep{arkhipova2017}.

According to the analysis of the CO lines, \ai presents a nebula (1.9\x\xd{-2}\msp) in which the expanding shell, with a velocity of $\sim$\,4\kmsp, would dominate the whole structure \citep{bujarrabal2013a,arkhipova2017}.

\ai shows a detached dust shell with $T\sim 200$\,K and it also presents TiO absorption bands that are formed in the upper cool layers of the extended atmosphere \citep{arkhipova2017}.
This source has been previously observed in radio lines. \citet{lintelhekkert1991} discovered the OH maser at 1\,612\,MHz in a survey of IRAS sources.
The \agua\ maser at 22\,235.08\,MHz was discovered by \citet{engelslewis1996}.
\citet{suarez2007,yoon2014} studied this source looking for SiO maser emission, but without success.
In view of the above results, \ai most likely is an O-rich source \citep[see also][]{arkhipova2017}.

\subsection{IRAS\,20056+1834}

This source has been studied in CO by \citet{bujarrabal2013a} and the analysis yields a nebular mass of \xd{-1}\msp, of which $\sim$\,22\% corresponds to the mass of the disk.
Apart from CO, there is no other molecular line detection in this source.

\subsection{\rsp} \label{rsct_res_previos}

The  bright RV\,Tauri variable R\,Sct  shows very irregular pulsations with variable amplitude \citep{kalaeehasazadeh2019} and a small IR excess \citep{kluska2019}, indicating that the SED is not clearly linked to the presence of a circumbinary disk. 
Nevertheless, the CO integrated flux of the innermost region of the nebula around \rs shows a  characteristic double peak, suggesting the presence of rotation.
Models for a disk with Keplerian dynamics and a very extended outflow are consistent with mm-wave interferometric observational data \citep[see][for more details]{gallardocava2021}. The nebular mass  is 3.2\x\xd{-2}\msp, where $\sim$\,75\% corresponds to the mass of the outflow.
The hypothesis of a rotating disk is reinforced by interferometric data in the \textit{H}-band showing a very compact ring \citep{kluska2019}.
The binarity of \rs is still questioned, but as Keplerian disks are only detected around binaries in the case of evolved stars, it could well be binary as well. 

In particular, R\,Sct presents composite CO line profiles including a narrow component, which very likely represents emission from the rotating disk \citep{bujarrabal2013a, gallardocava2021}.
The chemistry of this source has been studied in IR by \citet{matsuura2002} and \citet{yamamura2003}, who find that the IR spectra of \rs is dominated by molecular emission features, especially from \agua. These molecules are probably located in a spherical extended atmosphere. In addition, SiO, CO$_{2}$, and CO bands have been identified.
\citet{lebregillet1991} detected photospheric absorption Fe\,\textsc{i} lines, with the double absorption and emission Ti\,\textsc{i} profiles observed, as well as TiO.

\section{Observations and data reduction}
\label{obs}

Our observations were performed using the 30\,m\,IRAM telescope (Granada, Spain) and the 40\,m\,Yebes telescope (Guadalajara, Spain). We observed at the 1.3, 2, 3, 7, and 13\,mm bands (see \tab\ref{rangos}).
Our observations required a total telescope time of $\sim$\,600\,hours distributed over two the telescopes and for several projects (see \tab\ref{rangos}).

The data reduction was carried out with the software CLASS\footnote{Continuum and Line Analysis Single-dish Software (CLASS) is part of the GILDAS software package} within the GILDAS\footnote{GILDAS is a software package focused in reducing and analysing mainly millimeter observations from single-dish and  interferometric telescopes. It is developed and maintained by IRAM, LAOG/Université de Grenoble, LAB/Observatoire de Bordeaux, and LERMA/Observatoire de Paris. For further details, see \url{https://www.iram.fr/IRAMFR/GILDAS}} software package. For each source, 
we applied the standard procedure of data reduction that consists of rejecting bad scans, averaging the  good ones, and subtracting a baseline of a first-order polynomial.

\subsection{30\,m\,IRAM telescope}

We observed at the 1.3, 2, and 3\,mm bands using the 30\,m\,IRAM telescope. Our observations were performed in four different projects. Observations of \onp, \rsp, and the \rr were obtained between 13 and 19 February 2019 under project 179-18 for 80\,hours. The data of project 055-19 were obtained between 29 June and 1 July 2019, when we observed the \rr for an additional 62\,hours.
\iras and IRAS\,20056+1834 were observed between 9 September and 14 September 2020 for 42\,hours within project 042-20. As a continuation of the last project, we also observed AI\,CMi and HD\,52961 between 24 and 29 November 2020 for 45\,hours.
Finally, observations of \acp, IRAS\,19157$-$0257, and IRAS\,18123+0511 were obtained between 21 and 27 July 2021, and 27 August 2021 under project E04-20 for 90\,hours.

We connected the Fast Fourier Transform Spectrometer (FTS) units to the EMIR receiver with a resolution of 200\,kHz per channel. The Half Power Beam Width (HPBW) of the 30\,m\,IRAM telescope is 11$''$, 17$''$, and 28$''$ at 224, 145, and 86\,GHz, respectively.
The sources were observed using the wobbler-switching mode, which provides flat baselines. The subreflector was shifted every 2\,s with a throw of $\pm$120$''$ in the azimuth direction.
We obtained spectra for the vertical and horizontal linear polarization receivers. Both polarizations were obtained simultaneously and we did not find significant differences in their relative calibration.

The calibration was derived using the chopper-wheel method observing the sky, and hot and cold loads. The procedure was repeated every 15\,$-$\,20\,minutes, depending on the weather conditions. The observed peak emission values have been re-scaled (if applicable) comparing with the intensity of calibration lines observed in NGC\,7027 and CW\,Leo (IRC+10216). The absolute scale accuracy is of the order of 10\%, 10\%, and 20\% in the 3, 2, and 1.3\,mm receives, respectively.

\subsection{40\,m\,Yebes telescope}

We observed at the 7 and 13\,mm bands using the 40\,m\,Yebes telescope, \textit{Q}- and \textit{K}-band, respectively. We performed our observations in the course of two projects. We observed all the sources of our sample at 7\,mm between 29 May and 7 June 2020 for 50\,hours under project 20A009. In the same project, we also observed at 13\,mm between 20 and 27 May 2020 for 15\,hours.
The data of project 20B006 were obtained in several epochs, between 6 and 13 June, 24 and 26 November 2020, and 14 and 17 January 2021. We observed AI\,CMi, IRAS\,20056+1834, the \rrp, \onp, \acp, HD\,52961 and \iras at 7\,mm for 70\,hours. We also observed IRAS\,20056 and HD\,52961 at 13\,mm for 12\,hours between 24 and 29 October 2020.

The received signal was detected using the Fast Fourier Transform Spectrometer (FFTS) backend units.
The bandwidth of the \textit{Q}-band is 18\,GHz, the spectral resolution is 38\,kHz, and the HPBW is 37\,$-$\,49$''$. In this band, our sources were observed using the position-switching method, always using a separation of around 300$''$ in the azimuthal direction, obtaining flat baselines. We obtained spectra for the vertical and horizontal linear polarization simultaneously and we do not find significantly differences in their flux density calibration; see \citet{tercerof2021} for further technical details.
The bandwidth of the \textit{K}-band is 100\,MHz, the spectral resolution is 6.1\,kHz, and the HPBW is 79$''$.
Our sources were observed using the position-switching method, using separation of about 600$''$ in the azimuthal direction, obtaining flat baselines.
The observations at 13\,mm were carried out with dual circular polarization with small calibrations differences of $\leq$\,20\%.
Pointing and focus were checked in both \textit{Q}- and \textit{K}-bands every  hour  through  pseudo-continuum  observations of the SiO \vuno\ \unocero\ and \agua\ maser emission, respectively, towards evolved stars close to our sources that show intense masers; see \citet{devicente2016} for details. The pointing errors were always within 5$''$\,$-$\,7$''$.

\begin{table*}[h]
\caption{Detected and tentatively detected lines in this work.}
\small
\tiny
%\centering
\vspace{-5mm}
\begin{center}

\begin{tabular*}{\textwidth}{@{\extracolsep{\fill\quad}}lll@{\hspace{2ex}}lccccccl}
\hline \hline
\noalign{\smallskip}
%&&&&&\multicolumn{4}{|c|}{}\\
 %\\[-2ex]

\multirow{2}{*}{Source} &\multirow{2}{*}{Molecule} & \multicolumn{2}{c}{\multirow{2}{*}{Transition}} & I\,(peak) & $\sigma$ & $\int I\,dV$ & $\sigma \left(\int I\,dV \right)$ & Sp.\,Res. & $V_{LSR}$ &  \multirow{2}{*}{Comments} \\
& &  & & [Jy] & [Jy] & [Jy\,km\,s$^{-1}$] & [Jy\,km\,s$^{-1}$] & [km\,s$^{-1}$]  & [km\,s$^{-1}$] & \\
\hline
\\[-2ex]

\ac     &   SiO             &        \vuno  &   \dosuno         &   1.7E-02         &   5.1E-03         &   2.1E-02         &   9.1E-03         &      0.68         &      0.41         &      Tentative        \\

\hline
\\[-2ex]

Red\,Rectangle &              \cdoop         &              \vcero         &             \dosuno         &   1.7E-01         &   2.4E-02         &   9.5E-01         &   6.5E-02         &      0.53         &     $-$0.64         &                       \\
&              \cdsop         &              \vcero         &             \dosuno         &   1.8E-01         &   2.4E-02         &   7.9E-01         &   4.8E-02         &      0.52         &     $-$0.59         &                       \\
&                 HCN         &              \vcero         &            \unocero         &   1.3E-02         &   3.9E-03         &   8.6E-02         &   2.8E-02         &      2.64         &      7.86         &      Tentative        \\    
&                     SO         &              \vcero                        &              \trsoc         &   3.5E-02         &   6.7E-03         &   3.3E-01         &   6.6E-02         &      4.26         &     $-$1.15         &             \\
&               \agua         &              \vcero         &             \tragua         &   5.3E-02         &   1.1E-02         &   6.2E-01         &   9.3E-02         &      3.29         &      8.27         &       Tentative                \\

\hline
\\[-2ex]

89\,Her &            \cdoop         &              \vcero         &             \dosuno         &   1.2E-01         &   1.8E-02         &   2.0E-01         &   2.0E-02         &      0.27         &     -8.36         &                       \\  
&               \cdsop         &              \vcero         &             \dosuno         &   1.3E-02         &   5.1E-03         &   6.1E-02         &   5.2E-02         &      4.17         &    -11.10         &      Tentative        \\        
&                 HCN         &              \vcero         &            \unocero         &   1.2E-02         &   2.8E-03         &   2.1E-01         &   2.5E-02         &      2.64         &    -14.76         &          Tentative             \\
   &                 SiS         &              \vcero    &        \cincocuatro         &   1.5E-02         &   3.8E-03         &   3.3E-03         &   6.7E-03         &      0.64         &    -11.57         &                       \\
 &                 CS         &              \vcero  &            \tresdos         &   4.9E-02         &   8.1E-03         &   9.7E-02         &   1.4E-02         &      0.40         &     -7.42         &                       \\

\hline
\\[-2ex]

AI\,CMi &                 SiO         &              \vcero         &            \unocero         &   2.0E-02         &   5.4E-03         &   7.5E-02         &   2.0E-02         &      1.05         &     29.11         &                       \\
&                             &                             &             \dosuno         &   1.1E-01         &   1.8E-02         &   8.8E-01         &   7.0E-02         &      0.67         &     29.02         &                       \\
 &                            &                             &        \cincocuatro         &   2.9E-01         &   3.7E-02         &   2.0E+00         &   1.0E-01         &      0.54         &     29.00         &                       \\
  &                           &               \vuno         &            \unocero         &   1.4E+00         &   9.2E-03         &   2.9E+00         &   1.8E-02         &      0.27         &     28.87         &                       \\
   &                          &                             &             \dosuno         &   1.9E+00         &   1.7E-02         &   4.7E+00         &   4.8E-02         &      0.68         &     28.81         &                       \\
    &                         &               \vdos         &            \unocero         &   4.3E-01         &   8.9E-03         &   6.4E-01         &   1.6E-02         &      0.27         &     29.85         &                       \\                   
     &          \vsio         &              \vcero         &            \unocero         &   7.7E-03         &   3.9E-03         &   4.9E-02         &   1.5E-02         &      2.13         &     35.04         &      Tentative        \\
       &                      &                             &            \dosuno         &   2.6E-02         &   7.5E-03         &   2.5E-01         &   5.7E-02         &      2.73         &     28.52         &         Tentative              \\
      &         \tsio         &              \vcero         &            \unocero         &   1.2E-02         &   3.2E-03         &   8.4E-02         &   1.7E-02         &      2.16         &     29.03         &          Tentative             \\
     &            SiS         &               \vuno         &          \ochosiete         &   1.9E-01         &   4.4E-02         &   1.9E-01         &   5.4E-02         &      0.41         &     32.36         &      Maser?        \\
       &           SO         &              \vcero         &              \trsou         &   3.8E-02         &   9.0E-03         &   3.1E-01         &   4.1E-02         &      1.36         &     28.86         &                       \\
        &                     &                             &              \trsoc         &   5.0E-01         &   3.5E-02         &   4.3E+00         &   1.1E-01         &      0.53         &     28.88         &                       \\
         &      \sod         &              \vcero                    &            \trsodos         &   1.8E-01         &   2.9E-02         &   1.1E+00         &   9.0E-02         &      0.80         &     28.68         &                       \\
         &      \agua         &              \vcero         &             \tragua         &   1.6E+00         &   5.4E-02         &   9.8E-01         &   5.8E-02         &      0.16         &     27.18         &                       \\

\hline
\\[-2ex]

IRAS\,20056+1834 &                SiO         &                       \vuno         &            \unocero         &   1.5E-02         &   3.2E-03         &   4.0E-02         &   6.7E-03         &      0.53         &    -22.40         &                       \\
  &                           &               \vdos         &            \unocero         &   2.9E-02         &   3.9E-03         &   1.1E-01         &   1.1E-02         &      0.53         &    -20.56         &                       \\

\hline
\\[-2ex]

R\,Sct &             SiO         &              \vcero         &            \unocero         &   7.9E-02         &   8.8E-03         &   0.29         &   2.1E-02         &      0.53         &     56.02         &                       \\
 &                            &                             &             \dosuno         &   3.5E-01         &   6.8E-03         &   1.6E+00         &   2.8E-02         &      0.67         &     57.02         &                       \\
 &                            &                             &        \cincocuatro         &   1.3E+00         &   5.6E-02         &   5.5E+00         &   1.4E-01         &      0.27         &     57.50         &                       \\      
 &                            &               \vuno         &            \unocero         &   9.8E-02         &   7.4E-03         &   0.21         &   1.7E-02         &      0.53         &   54.32         &           \\
  &                           &                             &             \dosuno         &   5.4E-01         &   7.3E-03         &   1.0E+00         &   2.2E-02         &      0.68         &     53.50         &                       \\                 
  &                           &               \vdos         &            \unocero         &   2.9E-01         &   4.4E-02         &   0.64         &   7.3E-02         &      0.53         &   55.32       &        \\                                                 
   &                          &              \vseis         &            \unocero         &   2.3E-02         &   6.7E-03         &   8.2E-02         &   1.6E-02         &      1.10         &     52.88         &      First \vseis\ detection        \\
  &             \vsio         &              \vcero         &            \unocero         &   2.9E-02         &   7.8E-03         &   2.6E-02         &   1.1E-02         &      0.53         &     56.31         &      Tentative        \\
   &                          &                             &            \dosuno         &   8.0E-02         &   6.3E-03         &   3.5E-01         &   2.0E-02         &      0.68         &     57.98         &                       \\
    &           \tsio         &              \vcero         &            \unocero         &   2.6E-02         &   7.7E-03         &   2.6E-02         &   1.2E-02         &      0.54         &   54.58         &   Tentative        \\
   &                          &                             &            \dosuno         &   6.1E-02         &   6.5E-03         &   2.7E-01         &   1.7E-02         &      0.69         &     57.35         &                       \\
   &            \hcom         &              \vcero         &            \unocero         &   4.2E-02         &   5.3E-03         &   1.8E-01         &   1.3E-02         &      0.66         &     56.68         &                       \\                                    
  &              SO         &              \vcero         &              \trsoc         &   3.0E-02         &   8.3E-03         &   1.9E-01         &   4.4E-02         &      2.13         &     56.94         &                       \\
     &          \agua         &              \vcero         &             \tragua         &   1.7E+00         &   3.0E-02         &   2.0E+00         &   4.7E-02         &      0.41         &     56.75         &                       \\
\hline
\end{tabular*}
% \\[1ex]

\end{center}
\small
\vspace{-1mm}
\textbf{Notes.} For complete tables including upper limits, see \app\ref{tablas_completas}.

\label{lineas_detectadas}
\end{table*}

\section{Observational results}
\label{resultados}

We compiled a list of the detected molecular transitions, together with upper limits for undetected lines, summarized in \app\ref{tablas_completas}. 
Additionally, we summarized the main results from detected and tentatively detected lines in \tab\ref{lineas_detectadas}. 
The tables list, for the main transitions, the peak intensity flux (I\,[Jy]), its associated noise ($\sigma$\,[Jy]), the integrated line intensity ($\int I\,dV$\,[Jy\,km\,s$^{-1}$]), its associated uncertainty ($\sigma \left(\int I\,dV \right)$\,[Jy\,km\,s$^{-1}$]), the spectral resolution (Sp.\,Res.\,[km\,s$^{-1}$]), and the velocity centroid of the line measured with respect to the local standard of rest ($V_{LSR}$\,[km\,s$^{-1}$]).

All our sources were previously detected in CO lines \citep{bujarrabal2013a}. We present the first detection of other mm-wave lines (see \tab\ref{lineas}) in these sources and this work is one of the few that systematically surveyed pPNe in the search for molecules.
Our detected lines at 1.3, 2, 3, 7, and 13\,mm (\tab\ref{lineas_detectadas}) provide new data for the ten binary \pagb stars of our survey (\tab\ref{prop}). 
We  detected new lines in the Red\,Rectangle, 89\,Her, AI\,CMi, IRAS\,20056+1834, and R\,Sct.
We also present significant upper limits for a large number of lines (see \app\ref{tablas_completas}), which often include valuable information.

\subsection{\acp} \label{resultados_acher}

We  observed this source at 1.3, 2, 7, and 13\,mm, resulting in no positive detections, attaining an rms of 15, 9, 7, and 36\,mJy, respectively (see \tab\ref{lineas_acher}). We also observed \ac at 3\,mm, where we  attained a rms value of 4\,mJy, and we tentatively detected the \vuno\ \dosuno\ SiO maser (see \fig\ref{fig:acher_mol_1}).

This tentative detection is centered at $\sim$\,0\kmsp, which is shifted respect to the CO line profiles \citep[][]{gallardocava2021,bujarrabal2013a}. We assume that the SiO maser emission would have come from the innermost region of the disk, which shows the highest velocity shifts, explaining the observed velocity.

\subsection{\rrp}

The \rr is the best-studied object of our sample \citep{bujarrabal2013a,bujarrabal2013b,bujarrabal2016}.
We present our results in \tabs\ref{lineas_rr} and \ref{lineas_detectadas}, and in \figs\ref{fig:rr_mol} and \ref{fig:rr_mol_ten} (and \ref{fig:rr_mol_carb}). We have detected the rarest CO isotopic species, namely \cdsop\ and \cdoop\ \dosuno. The values for the line peak and integrated intensity of the two lines are similar.
We have tentatively detected HCN\,\unocero\, with signal-to-noise ratio (S/N\,$>$\,3). This detection is consistent with the presence of H$^{13}$CN\,$J=4-3$ \citep{bujarrabal2016}. The presence of these molecules, together with the \textsc{C\,i}, \textsc{C\,ii} lines, and PAHs, could be due to the development of a PDR in dense disk regions close to the central stellar binary system. The origin of HCN could be photoinduced chemistry, because it has been detected in the innermost regions of several disks in young stars. This fact could be also present in relatively evolved PNe with UV excess \citep{bublitz2019}, and is predicted by theoretical modelling of the PDR chemistry \citep[e.g.,][]{agundez2008}.

The line profile of these transitions shows the double peak characteristic of rotating disks \citep{guilloteaudutrey1998, bujarrabal2013a, guilloteau2013}. It implies that the emission of these molecules comes from the innermost region of the rotating component of the nebula. 

In addition, we detected SO $J_{N}=6_{6}-5_{4}$. The line profile of the SO line is narrow, as the CO lines, which suggests that oxygen rich material could be located in the Keplerian disk. We also tentatively detected emission of \agua\ \tragua. The centroid of this line is 8.3\kmsp, and this shift in velocity could have its origin in maser emission. We have confirmed the detection of \agua\ maser emission at 325\,GHz in our new ALMA maps (private communication).
We  also improved the rms at 90\,GHz given by \citet{liujiang2017} by a factor of 13 (see \tab\ref{lineas_rr}).

\subsection{\onp}

We present our results in \tabs\ref{lineas_89her} and \ref{lineas_detectadas}, and in \figs\ref{fig:89her_mol} and \ref{fig:89her_mol_ten}. 
We  detected \cdoop\ \dosuno\ and tentatively \cdsop\ \dosuno.
We  also detected CS \tresdos\ and SiS \cincocuatro. All these lines show narrow profiles.
Additionally, we tentatively detected HCN \unocero. 
Based on the shape of the observed profiles, we are confident that the emission of these molecules comes from the Keplerian disk and not from the extended outflow.

\subsection{HD\,52961}

We did not detect any emission at the observed frequencies, but we highlight that we present significant upper limits (see \tab\ref{lineas_hd52961}). We attained a rms of 40, 45, and 18\,mJy at 1.3, 2, and 3\,mm, respectively. The observation at 7 and 13\,mm was focused in the detection of SiO and \agua\ maser emission and we attained a rms of 6 and 31.5\,mJy, respectively.

\subsection{IRAS\,19157$-$0257 and IRAS\,18123+0511} \label{lasiras}

We focus our mm-wave single-dish observations in the detection of SiO and \agua\ maser emission at 7 and 13\,mm.
We have not confirmed the lines in IRAS\,19157$-$0257 and IRAS\,18123+0511, but we provide upper limits (we present our results in \tabs\ref{lineas_iras19157} and \ref{lineas_iras18123}). Some of them clearly improve the rms achieved by previous works by other authors.

We attained an rms in IRAS\,19157$-$0257 of 50, 40, 14 and 34\,mJy at 2, 3, 7, and 13\,mm, respectively.
We attain an rms in IRAS\,18123+0511 of 90, 55, 30, 17 and 24\,mJy at 1.3, 2, 3, 7, and 13\,mm, respectively.

\subsection{\irasp}

We did not detect any emission in \iras at the observed frequencies, but we highlight that we present significant upper limits (see \tab\ref{lineas_iras19125}). We have attained a rms of 50, 20, 10, and 23\,mJy at 1.3, 3, 7, and 13\,mm, respectively.

\subsection{\aip}
\label{resultados_aicmi}

We present our results in \tabs\ref{lineas_aicmi} and \ref{lineas_detectadas}, and in \figs\ref{fig:aicmi_mol_1_ten}, \ref{fig:aicmi_mol_1}, and \ref{fig:aicmi_mol_2}. 
We detected several lines of thermal SiO: \unocero, \dosuno, and \cincocuatro. We also detected SiO isotopic species, \vsio\ \unocero\ and \dosuno\ and \tsio\ \unocero.
We find very intense SiO maser emission for transitions \vuno\ \unocero\ and \dosuno\, and \vdos\ \unocero.
We find wide composite profiles of SO $J_{N}=2_{2}-1_{1}$ and $J_{N}=6_{5}-5_{4}$ and SO$_2$ \trsodos.
In addition, we detected \agua\ \tragua\ (this line was previously detected, see \sect\ref{aicmi_res_previos}).
We detected SiS\,\vuno\,\ochosiete, which, if confirmed, will also be the first detection of this species in \aip. However, we note that we did not detect any SiS $v=0$ in the observed bands, which is quite surprising but not impossible (SiS $v=1$ emission could be due to some weak maser emission).

All the detected thermal lines show composite profiles with a narrow component, in a similar way to what we find in the CO line profiles of this source and in \rsp. We think that both sources could be very similar.

\subsection{IRAS\,20056+1834} \label{resultados_iras20056}

We observed this source in 1.3 and 3\,mm, resulting in no positive detections and we attained rms values of 45 and 18\,mJy, respectively. We also observed this source at 7\,mm, where we detected maser emission, SiO\,\vuno\,\unocero\ and \vdos\,\unocero\ (see \fig\ref{fig:iras20056_mol}). There are no signs of \agua\ maser emission at 13\,mm, but we provide a rms of 27\,mJy (see \tab\ref{lineas_iras20056} for further details).
 
The SiO maser emission is centered at $\sim$\,$-$20\kmsp, which is shifted with respect to the CO line profiles by $\sim$\,10\kmsp. This fact is supported by \citet{klochkova2007}, who highlight that velocities measured from lines formed in the photosphere are variable. They reveal differential line shifts up 10\kmsp. Shifts between thermal and maser lines are in any case not usual.

We think that IRAS\,20056+1834 could be similar to \rs and \ai (a binary system surrounded by a Keplerian disk and an extended outflow that dominates the whole nebula), which would explain the chemical similarities between these sources.

\subsection{\rsp}
\label{resultados_rsct}

We present our results in \tabs\ref{lineas_rsct} and \ref{lineas_detectadas}, and in \figs \ref{fig:rsct_mol_1_ten}, \ref{fig:rsct_mol_1}, and \ref{fig:rsct_mol_2}. 
We detected several lines of thermal SiO: \unocero, \dosuno, and \cincocuatro. We also detected SiO isotopic species, \vsio\ and \tsio\ \unocero\ and \dosuno.
All of them show a narrow profile, which is coincident with the narrow component of the composited CO line profiles.
We also detected SiO maser emission for transitions \vuno\ \unocero\ and \dosuno, \vdos\ \unocero, and the first-ever detection of the \vseis\ \unocero. 
We detected SO \trsoc, and the \agua\ \tragua\ maser emission (see \sect\ref{rsct_res_previos}).

In addition, we also detected \hcom \unocero. This molecule presents weak emission in AGB stars when detected \citep{bujarrabal1994a,pulliam2011}. Furthermore, \hcom\ is detected in other standard pPNe, such as CRL\,618 \citep{sanchezcontreras2004}, OH\,231.8+4.2 \citep{sanchezcontreras2015}, or M1$-$92 \citep{alcolea2019}.
The detection of HCO$ ^{+}$ in the envelope of \rs implies that the disk is (partially) ionized. 
The most probable reason for the presence of \hcom\ is that the central star, which is now in the \pagb phase, can partially ionize the closest and densest regions. These regions correspond to the disk, which would justify the narrow profile of this line (see \fig\ref{fig:rsct_mol_1}).
Another way to explain the presence of ionized gas could be the presence of shocks, in which the abundance of SO is also favored.

\section{Discussion}
\label{discusiones}

\begin{table}[h]
\caption{Chemistry of the envelopes around binary \pagb stars observed in this paper.}
\small
%\tiny
%\centering
\vspace{-5mm}
\begin{center}
%\resizebox{\columnwidth}{!}{
\begin{tabular}{l@{\hskip 0.2cm}l@{\hskip 0.2cm}l}
\hline \hline
\noalign{\smallskip}
%&&&&&\multicolumn{4}{|c|}{}\\
 %\\[-2ex]

Source & Chemistry   & Comments  \\

\hline
\\[-2ex]
\vspace{1mm} 

\acp &  O-rich &  Tentative SiO maser emission \\
\vspace{1mm} 
Red\,Rectangle &  O-rich &  See \fig\ref{fig:cuadros_mol_mol} and \sect\ref{rr_corich}  \\
\vspace{1mm}  
89\,Herculis   &   C-rich & See \fig\ref{fig:cuadros_mol_mol} \\
\vspace{1mm} 
HD\,52961    & O-rich & See \citet{gielen2011b} \\

AI\,CMi &  O-rich & See \fig\ref{fig:cuadros_mol_mol} \\
&& SiO, \agua\, and OH maser emission \\
\vspace{1mm} 
&& See \citet{arkhipova2017} \\
\vspace{1mm} 
IRAS\,20056+1834 &  O-rich & SiO maser emission \\

R\,Scuti &   O-rich & See \fig\ref{fig:cuadros_mol_mol}  \\
&&  SiO and \agua\ maser emission \\
&&  \agua\ emission in IR spectra \\

\hline
\end{tabular}
%}
% \\[1ex]

\end{center}
\small
\vspace{-1mm}
\textbf{Notes.} We did not detect any thermal emission of O-bearing molecules in \ac and IRAS\,20056+1834, but we  detected SiO maser emission (see \sects\ref{resultados_acher} and \ref{resultados_iras20056}). HD\,52961 is cataloged as O-rich, based on discussions by \citet{gielen2011b}.

\label{ocrich}
\end{table}

\subsection{Molecular richness} \label{pobreza_molecular}

Prior to this work, the presence of molecules other than CO in these objects was practically unknown. We compared our results with standard AGB stars \citep[see e.g.,][]{bujarrabal1994a}, which are the precursors of our objects, representing a homogeneous group that has been studied in great detail and whose envelopes are rich in molecular emission. Additionally, photodissociation effects in AGB envelopes only appear in outer layers.
The comparison of our results with PNe is more difficult, because these objects present large differences in the molecular gas content \citep[][]{huggins1996,santandergarcia2021}, which is also present in young PNe \citep{bujarrabal1988,bujarrabal1992}. This fact is most likely due to molecular photodissociation, which strongly depends on the different evolutionary state of each source \citep{bachiller1997a, bachiller1997b, bublitz2019,ziurys2020}.
An added difficulty is that there is no complete survey of PNe or young PNe to compare with.

\begin{figure*}[h]
\center
\includegraphics[width=\sz\linewidth]{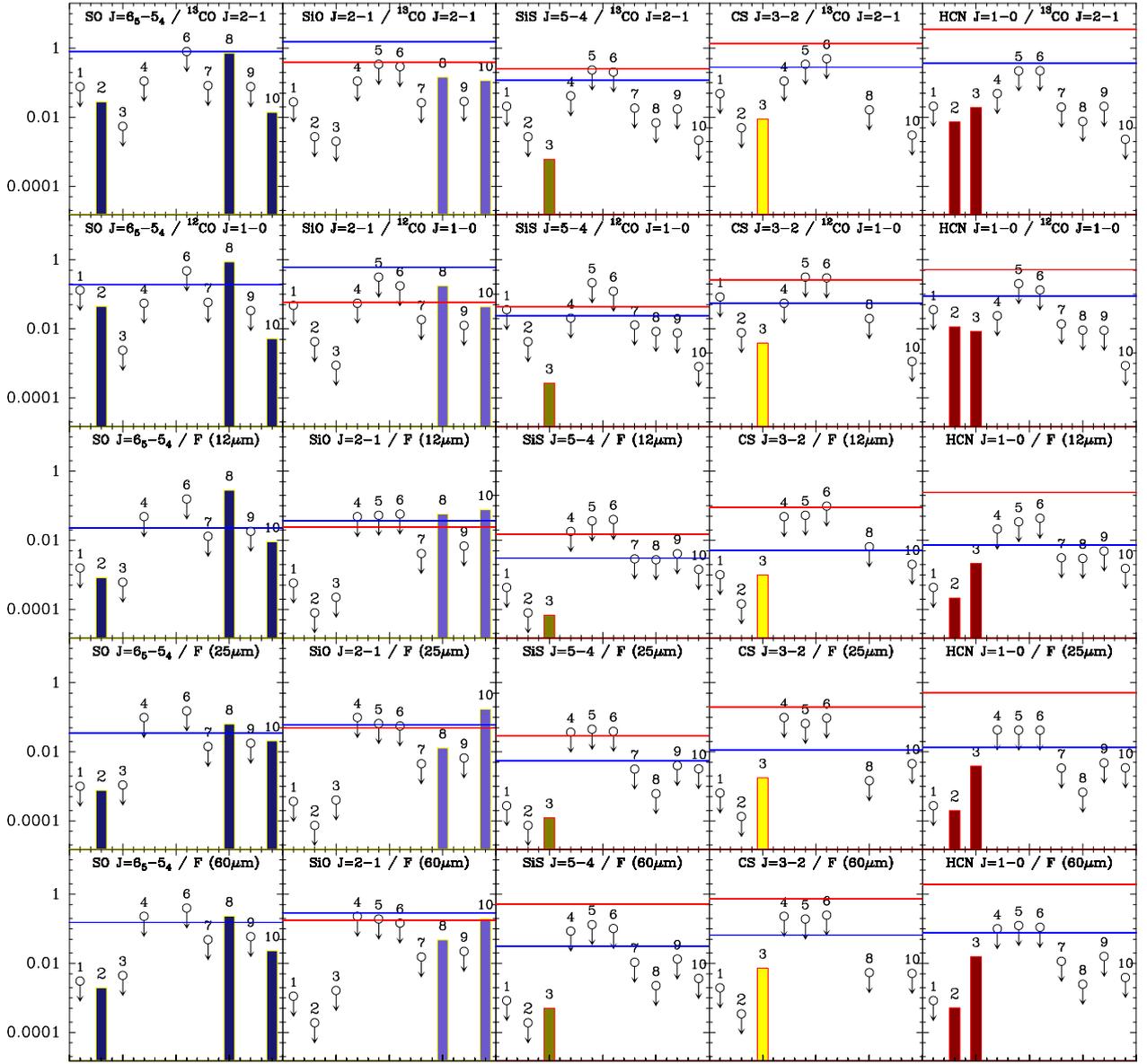}
\caption{\small Ratios of integrated intensities of molecules (SO, SiO, SiS, CS, and HCN) and infrared emission (12, 25, and 60\,$\mu$m) in binary \pagb stars. 
Upper limits are represented with empty circles. Blue and red lines represent the averaged values for O- and C-rich AGB CSEs. Sources are ordered by increasing outflow dominance and enumerated as follows: 1\,$-$\,\acp, 2\,$-$\,\rrp, 3\,$-$\,\onp, 4\,$-$\,HD\,52961, 5\,$-$\,IRAS\,19157$-$0257, 6\,$-$\,IRAS\,18123+0511, 7\,$-$\,\irasp, 8\,$-$\,\aip, 9\,$-$\,IRAS\,20056+1834, and 10\,$-$\,\rsp. Sources 1 and 2 are disk-dominated binary \pagb stars, sources 6 to 10 are outflow-dominated, while sources 3 and 4, and 5 are intermediate cases. We note the broad range (on a logarithmic scale) of intensity ratios. We always find uncertainties lower than $\sim$\,20\% in these ratios, which are basically dominated by that of the integrated intensity of the molecules other than CO (see main text for details).}
    \label{fig:moleculas_raras}  
\end{figure*}

In \fig\ref{fig:moleculas_raras}, we show integrated intensity ratios between the main molecules (SO, SiO, SiS, CS, and HCN) and CO (\trece\dosuno\ and with \doce\unocero). Additionally, we compare these molecular integrated intensities with infrared emission at 12, 25, and 60\,$\mu$m \citep[from the \textit{IRAS} mission, see][]{neugebauer1984}. We used these three IRAS wavelengths because the shape of the SEDs seems to be strongly related with the evolutionary state of the source \citep[see \fig\ref{fig:diagramacolorcolor} and][]{kwok1989,hrivnak1989,vanderveen1995,oomen2018}.
Molecular emission in AGB stars is usually compared to 60\,$\mu$m, since this band adequately represents the bulk of the circumstellar emission. 
Many pPNe present a bimodal SED with an excess at 25\,$-$\,60\,$\mu$m and less emission at 5\,$-$\,10\,$\mu$m, because of a lack of dust characterized by an intermediate temperature.
On the contrary, our objects tend to present a significant NIR excess at 12 and 25\,$\mu$m, and these two bands are probably better tracing most of the nebular material. However, since we want to study the gast-to-dust ratios in our sources and compare them with those in other objects, we needed to use all these three CO-to-IRAS ratios.
We always find uncertainties less than $\sim$\,20\% in our ratios, which basically are dominated by that of the integrated intensity of the molecules other than CO (see tables of  \app\ref{tablas_completas} for further details). These uncertainties are too small to be represented in the figures; thus, we note the very broad range (on a logarithmic scale) of the intensity ratios.
Our results were compared with  molecular emission of AGB stars. Blue and red horizontal lines represent averaged values of the molecular emission for O- and C-rich AGB stars, respectively.
We note that the emission of molecules other than CO in our sources is low. This fact is especially remarkable in those sources which are dominated by their rotating disk (\ac and the \rrp) or intermediate cases (such as \onp).

\begin{figure*}[!h]
\center
\includegraphics[width=1 \linewidth]{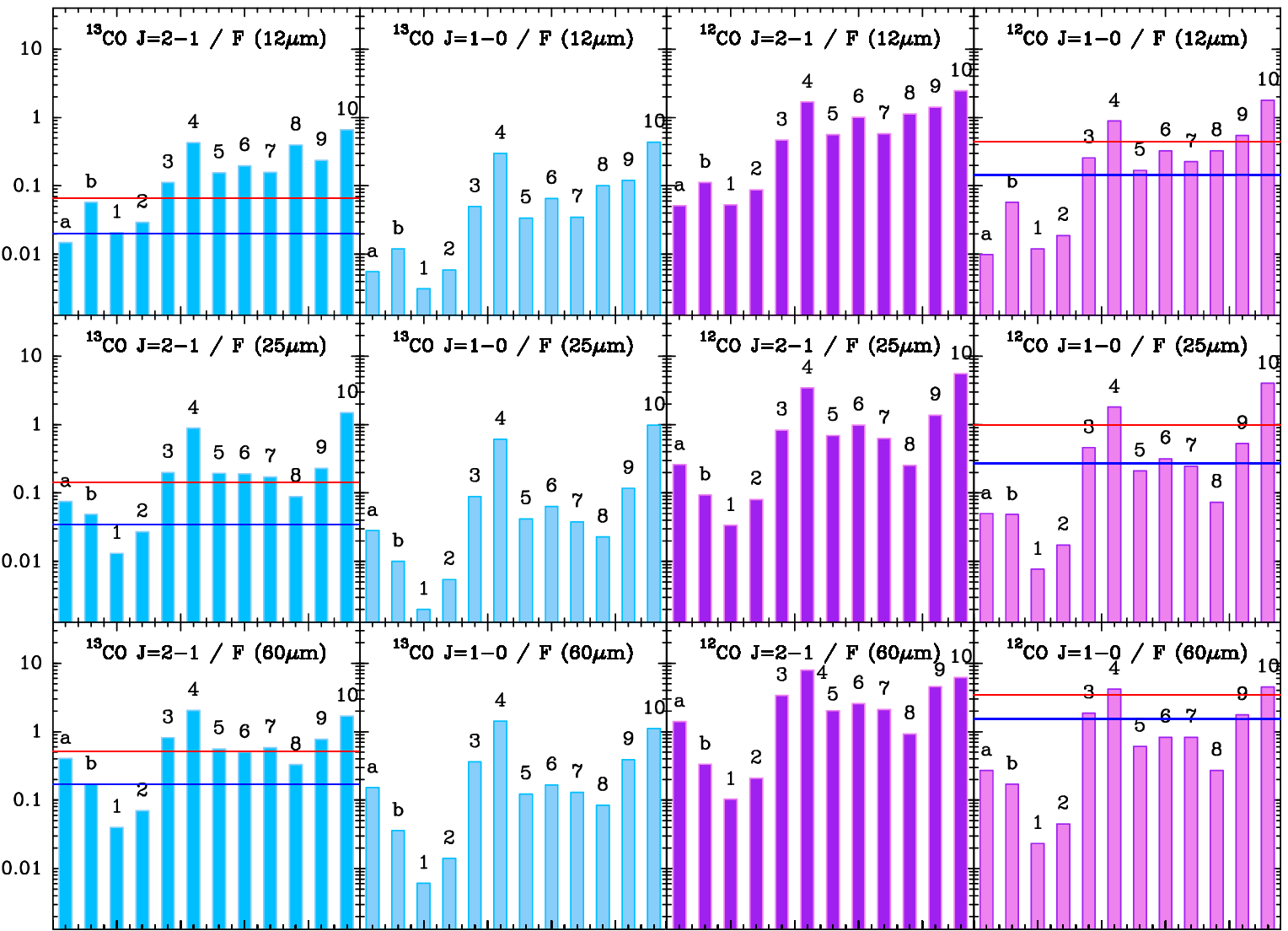}
\caption{\small 
Ratios of integrated intensities of \doce (blue) and \trece (purple) \dosuno\ and \unocero\ and infrared emission (12, 25, and 60\,$\mu$m) in binary \pagb stars. Blue and red lines represent the averaged values for O- and C-rich AGB CSEs. Sources are are ordered by increasing outflow dominance and enumerated as follows: a\,$-$\,HR\,4049, b\,$-$\,DY\,Ori, 1\,$-$\,\acp, 2\,$-$\,\rrp, 3\,$-$\,\onp, 4\,$-$\,HD\,52961, 5\,$-$\,IRAS\,19157$-$0257, 6\,$-$\,IRAS\,18123+0511, 7\,$-$\,\irasp, 8\,$-$\,\aip, 9\,$-$\,IRAS\,20056+1834, and 10\,$-$\,\rsp. Souces a and b are not included in this mm-wave survey. Sources 1, 2, a, and b are disk-dominated binary \pagb stars, sources 6 to 10 are outflow-dominated, and sources 3, 4, and 5 are intermediate cases. We note the broad range (on a logarithmic scale) of intensity ratios.
We find uncertainties less than $\sim$\,10\% in these ratios, which are basically dominated by that of the integrated intensity of CO \citep[see][]{bujarrabal2013a}.
}
    \label{fig:moleculas_co}  
\end{figure*}

In \fig\ref{fig:moleculas_co}, we show, $^{12}$CO\,-\,to\,-\,IR and $^{13}$CO\,-\,to\,-\,IR ratios for the integrated intensity of \dosuno\ and \unocero. 
For transitions compared with the IR fluxes, the \doce\ emission is relatively low with respect to the levels exhibited by AGB stars. 
Again, we find two subclasses in our sample. Those sources in which the Keplerian disk dominates the nebula tend to present a lower relative intensity of \doce\ compared to the outflow-dominated ones.
On the contrary, we do not see this effect in \trece\ emission, which seems to be relatively intense compared with \docep and comparable to that from AGB stars.
The difference found in this ratio in the disk- and the outflow-dominated sources could be due to higher optical depths in the \doce\dosuno\ line. The emission of the outflow-dominated sources comes from relatively extensive and very diffuse areas, thus they are expected to show low optical depths in the \doce\dosuno\ line. For example, in the case of \rsp, the emitting area exceeds 2\x\xd{17}\,cm. However, in the case of the very well known disk-dominated sources, such as \acp, the emitting area does not exceed 2\x\xd{16}\,cm, which means that it is ten times smaller than in the case of \rsp. Thus, based on this discussion and the results (see \fig\ref{fig:moleculas_raras}), we find a real difference in the relative abundances that seems to confirm that the outflow-dominated sources present relative higher \trece\ abundance. This fact was previously noted in \rsp, which is an outflow-dominated source \citep[][]{gallardocava2021,bujarrabal1990}, where a very low $^{12}$C\,/\,$^{13}$C ratio was reported.

\subsection{Discrimination between O- and C-rich envelopes} \label{corich}

\begin{figure*}[h]
\center
\includegraphics[width=1 \linewidth]{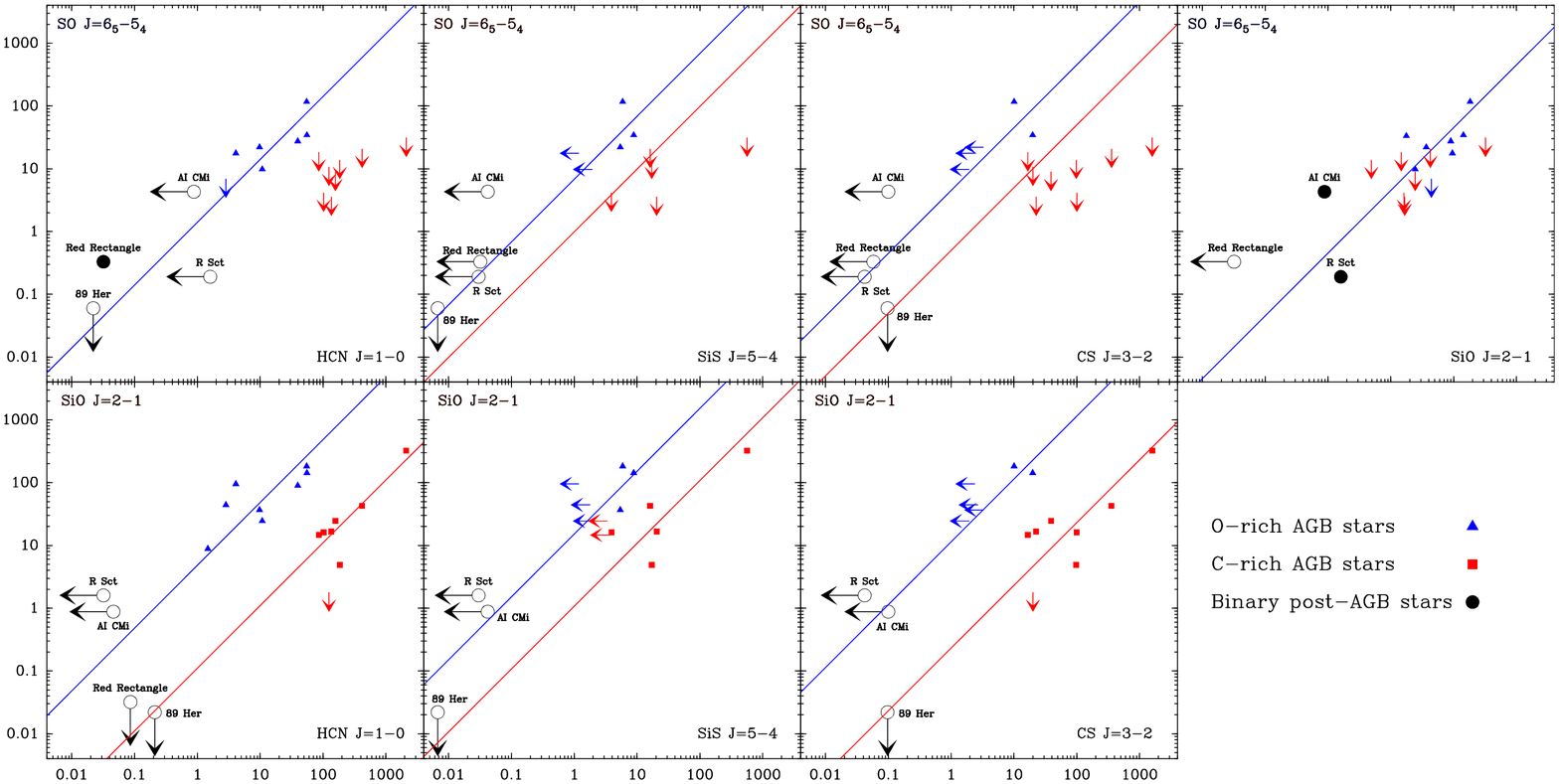}
\caption{\small Integrated intensities of pairs of molecular transitions in binary \pagb stars (black), as well in O- and C-rich stars (blue and red). Blue and red lines represent the averaged values for O- and C-rich AGB CSEs (only in the case of detections). The X and Y axes represent the integrated intensities of the observed transitions in logarithmic scale and units of Jy\,\kmsp. The upper limits are represented bv arrows.}
    \label{fig:cuadros_mol_mol}  
\end{figure*}

In general, evolved stars can be classified as O- and C-rich according to their relative O\,/\,C elemental abundance. It is known that this difference, even if the O\,/\,C abundance ratio is not always  shown to vary much from 1, has important effects on the molecular abundances.
SiO, SO are O-bearing molecules and their lines are much more intense in O-rich envelopes than in the C-rich ones, while HCN and CS (HC$_{3}$N and HNC) are C-bearing molecules (but also SiS), whose lines are more intense in C-rich envelopes than in O-rich \citep[see e.g.][]{bujarrabal1994a, bujarrabal1994b}. Additionally, SiO maser emission is detected in M- and S-type AGB stars, and \agua\ maser emission is only seen in M-type stars. In any case, SiO and \agua\ maser emission is seen exclusively towards O-rich envelopes \citep{kim2019}.
The analysis of integrated intensities of pairs of molecular transitions is very useful to distinguish between O- and C-rich environments (see \fig\ref{fig:cuadros_mol_mol}). Integrated intensities are larger for O-rich than for C-rich objects when an O-bearing molecule is compared with a C-bearing one.
Our results are compared with CSEs around AGB stars, which are deeply studied objects and they are prototypical of environments rich in molecules \citep[for this comparison, AGB data is taken from][which represent a wide sample of CSEs around evolved stars in the search molecules other than CO]{bujarrabal1994a, bujarrabal1994b}.
In \fig\ref{fig:cuadros_mol_mol}, blue and red lines represent averaged values for these ratios in O- and C-rich AGB CSEs, respectively.

In the case of \ai and \rsp, the integrated intensities of SO and SiO compared with C-bearing molecules are remarkably larger than for C-rich evolved stars. Both sources present SiO and \agua\ maser emission \citep[\ai also presents OH maser emission according to][]{lintelhekkert1991}. 
Additionally, the IR spectra of \rs is mainly dominated by \agua\ emission \citep{matsuura2002,yamamura2003}. These facts lead us to classify \ai and \rs as O-rich.
The \rr presents O- and C-bearing molecular emission. This case will be examined in more detail in \sect\ref{rr_corich}, but according to \fig\ref{fig:cuadros_mol_mol} and the tentative \agua\ maser detection, we catalog the \rr as O-rich too.
There are no signs of O-bearing molecules in \on and we detected HCN, CS, and SiS, so the chemistry of this source is absolutely compatible with a C-rich nebula.
Finally, we did not detect any thermal SiO emission in IRAS\,20056+1834, but we have clearly confirmed the existence of SiO maser emission and tentative SiO maser emission in \acp (see \sects \ref{resultados_acher} and \ref{resultados_iras20056}), which is exclusive of O-rich environments. 

Based on the integrated intensities ratios in \fig\ref{fig:cuadros_mol_mol} and the maser detection of O-bearing molecules, we can classify some of our sources as O-\,/\,C-rich (see a summary of our results in \tab\ref{ocrich}). 
\on presents a O\,/\,C\,<\,1 chemistry. On the contrary, \acp, \aip, IRAS\,20056+1834, and \rs present a O\,/\,C\,>\,1 chemistry. 
The chemistry of the \rr could be uncertain, but we also classified this source as O-rich (see \sect\ref{rr_corich} for further details).
In any case, we recall that our sources are still poorly studied and that additional work, both observational and theoretical, is needed to confirm our findings on the general chemistry of these sources.

\begin{figure}[h]
\center
\includegraphics[width=1 \linewidth]{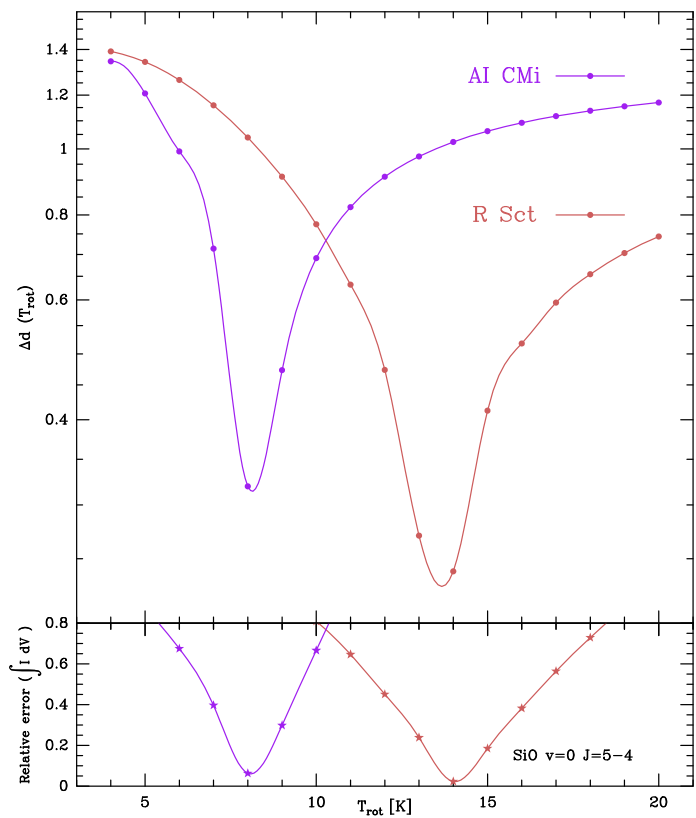}
\caption{\small \textit{Top}: Variations of the relative difference of the abundances produced by the line-ratio method as a function of the rotational temperature $\Delta d\,(T_{rot})$ in the case of \ai and \rsp. 
The minimum relative difference corresponds to the best-fit value of $T_{rot}$.
This procedure can only be performed for those sources in which several transitions of the same molecule have been observed. See text and \app\ref{ap_abundancias} for more details.
\textit{Bottom}: Variations of the relative error of the observational integrated intensity line and the predicted one as a function of the rotational temperature for SiO $J=5-4$. The predicted integrated intensity lines are derived from \eq\ref{eq:tmb_masa} using the estimated abundance and rotational temperature.}
    \label{fig:abundancia_media}  
\end{figure}

\begin{table}[h]
\caption{Abundances estimated.}
%\small
\tiny
%\centering
\vspace{-5mm}
\begin{center}
%\resizebox{\columnwidth}{!}{
\begin{tabular}{l@{\hskip 0.2cm}l@{\hskip 0.2cm}c@{\hskip 0.2cm}c@{\hskip 0.2cm}c}

\hline \hline
\noalign{\smallskip}
%&&&&&\multicolumn{4}{|c|}{}\\
 %\\[-2ex]
\multirow{2}{*}{Source} & \multirow{2}{*}{Molecule} & $T_{rot}$ & \multirow{2}{*}{$\langle X \rangle$} &  \multirow{2}{*}{$\langle X_{c} \rangle$}\\
 & & [K] & \\

\hline
\\[-2ex]
%\vspace{1mm} 

AI\,CMi       & SiO          &  8        &   1.3\x\xd{-8}  $^{*}$           &  3.1\x\xd{-8}  \\
              & $^{29}$SiO   &  8        &   3.7\x\xd{-9}   $^{*}$          &  9.3\x\xd{-8}  \\
\vspace{1mm} 
              & $^{30}$SiO   &  8        &    <\,1.1\x\xd{-8}  $^{*}$       &  <\,2.7\x\xd{-8}  \\

R\,Scuti      & SiO          &  14       &   5.6\x\xd{-9}           &  1.5\x\xd{-8}  $^{*}$  \\
              & $^{29}$SiO   &  14       &   5.0\x\xd{-10}          &  4.2\x\xd{-9}  $^{*}$  \\
              & $^{30}$SiO   &  14       &   1.1\x\xd{-9}           &  2.7\x\xd{-9}   $^{*}$ \\
\vspace{1mm}               
              & HCO$^{+}$    &  14       &   3.8\x\xd{-10}          & 1.0\x\xd{-9}  $^{*}$   \\

 \rr           & HCN          &  8\,$-$\,14  & (1.4\mm 0.5)\x\xd{-9}  $^{*}$  & $-$ \\
%\vspace{1mm} 
89\,Herculis  & HCN          &  8\,$-$\,14  & (1.1\mm 0.2)\x\xd{-10}  $^{*}$  & (1.9\mm 0.2)\x\xd{-9} \\
              & SiS          &  8\,$-$\,14  & (3.7\mm 0.5)\x\xd{-11}  $^{*}$  & (6.3\mm 0.8)\x\xd{-11} \\

              & CS           &  8\,$-$\,14  & (5.9\mm 0.8)\x\xd{-10}  $^{*}$  & (5.1\mm 3.9)\x\xd{-10} \\

%\rr           & HCN          &  8\,$-$\,14  & (0.9\,$-$\,1.9)\x\xd{-9}  & $-$ \\
%%\vspace{1mm} 
%89\,Herculis  & HCN          &  8\,$-$\,14  & (0.9\,$-$\,1.2)\x\xd{-10}  & (1.7\,$-$\,2.1)\x\xd{-9} \\
%              & SiS          &  8\,$-$\,14  & (4.1\,$-$\,3.2)\x\xd{-11}  & (7.0\,$-$\,5.5)\x\xd{-11} \\
%\vspace{1mm}            
%              & CS           &  8\,$-$\,14  & (6.7\,$-$\,5.1)\x\xd{-10}  & (1.2\,$-$\,8.9)\x\xd{-10} \\

\hline
\end{tabular}
%}
% \\[1ex]

\end{center}
\small
\vspace{-1mm}
\textbf{Notes.} Averaged abundances of the emitting gas for SiO, $^{29}$SiO, and $^{30}$SiO in the case of \ai and \rsp. 
$\langle X \rangle$ is the abundance considering the total mass of the envelope.
$\langle X_{c} \rangle$ is the abundance considering only the central region of the envelope and it represents $\sim$\,40\%, $\sim$\,40\%, $\sim$\,100\%, and $\sim$\,60\% of the total mass in the case of \aip, \rsp, the \rrp, and \onp, respectively. The asterisk, $^{*}$, represents our preference for each case.
\label{abun}
\end{table}

\subsection{Abundance estimates} \label{abundancias}

The methodology to estimate abundances is similar to that used in \citet{bujarrabal2001, quintanalacaci2007,bujarrabal2013a} and is described in \app\ref{ap_abundancias} in detail.
This procedure assumes optically thin emission, which is justified in our case, because the molecular lines present very weak emission (see \sect\ref{pobreza_molecular}).
To estimate the characteristic rotational temperature ($T_{rot}$), it is necessary to observe (at least) two transitions of the same molecule. We apply this procedure, using \eq\ref{eq:tmb_masa}, to \ai and \rs for SiO and its rare isotopic substitutions. We assume nebular masses derived from previous works \citep[see \tab\ref{prop},][]{bujarrabal2013a, gallardocava2021}.

After computing the abundances, we checked that our optically thin assumption is valid. We estimated the line opacities expected for the derived values of abundances and $T_{rot}$. These values are shown in \tab\ref{tau_calc}. We can see that in all cases, $\tau \ll 1$.

We estimate the rotational temperature ($T_{rot}$) from the observed line ratios. Different values of the abundance are calculated for each transition as a function of $T_{rot}$ (see \app\ref{ap_abundancias} for further details). The minimum relative difference of the abundances yields the best-fit rotational temperature (see \fig\ref{fig:abundancia_media} and \eqs\ref{eq:cantidad_d} and \ref{eq:error}).
For SiO, the best-fit $T_{rot}$ is 8 and 14\,K in \ai and \rsp, respectively. We find relative errors lower than 5\% between the integrated line intensity and the predicted one derived from \eq\ref{eq:tmb_masa} (see \fig\ref{fig:abundancia_media} \textit{bottom}). 
This relative error between the integrated intensity line and the predicted one considerably increases if we consider small variations with respect to our best-fit $T_{rot}$.
We can assume the rotational temperature of SiO for $^{29}$SiO and $^{30}$SiO, as well as for HCO$^{+}$, because the lines show similar excitation properties and line profiles (see \fig\ref{fig:rsct_mol_1}).

We see that most of the molecular line shapes are very similar to that of the CO lines (except the masers, whose lines have not been considered for the calculation of abundances; see \sect\ref{resultados}). We estimated the molecular mass of the envelopes of these sources based on the CO lines. Therefore, we can calculate relative abundances of molecules other than CO assuming that their emission comes from the same molecule-rich zone for which mass values were derived.
This is not the case of \rsp, whose thermal emission of molecules is narrower than that of CO. We think that the emission of these molecules very probably comes from the rotating disk and the high density region of the outflow (see \figs\ref{fig:rsct_mol_1_ten} and \ref{fig:rsct_mol_1}). We present abundance calculations for the total nebular mass, $\langle X \rangle$, and for that of the central region of the nebula, $\langle X_{c} \rangle$, which represents the rotating gas and the innermost regions of the outflow. 
In other words, $\langle X \rangle$ represents the relative abundance considering that the species are located in the whole nebula, while  $\langle X_{c} \rangle$ is the value obtained  assuming that the species are located  only in the densest part of the nebula.
According to our previous reasoning, for \rs our best option is to compute relative abundances with respect to the CO mass in these central regions, which in this case account for 40\% of the total nebular mass.
The shapes of the SiO, SO, and SO$_{2}$ line profiles of \ai are very similar to that in CO (see \fig\ref{fig:aicmi_mol_1}); therefore we think that the molecular emission of \ai must arise from the entire nebula.
We recall that in any case, these estimates are uncertain by a factor of 2 (in addition to uncertainties for other reasons) in view of the assumed fraction of involved material.
To estimate relative abundances assuming other fractions of the mass of the molecular rich component, $f$, with respect to the total nebular mass, we only need to multiply $\langle X \rangle$ by $f^{-1}$.

In the case of the \rr and \on there is only one detected transition per molecule, so we can only make very crude estimations of the abundance by assuming the obtained $T_{rot}$ of SiO for \ai and \rs and the nebular masses derived from previous works \citep{bujarrabal2016, gallardocava2021}.
We also estimated abundances for the whole nebula and for the inner regions: in the case of the \rr $\sim$\,90\% is located in the disk; in the case of \onp, 60\% of the total molecular mass corresponds to the rotating disk and the inner regions of the outflow. Based on the same reasoning as in the case of \aip, our best option for \rr and \on is to consider the abundance corresponding to the total molecular mass.

The calculated abundances for \aip, \rsp, \rrp, and \on can be seen in \tab\ref{abun} (we also show estimated optical depths in \tab\ref{tau_calc}).
We note the very low abundances we deduce, compared with standard AGB stars, which reinforces our conclusion, set out in \sect\ref{pobreza_molecular}. We can see again how the presence of molecules in disk-dominated or intermediate sources is much weaker than that in outflow-dominated ones.

\subsection{Dual nature of the \rr}
\label{rr_corich}

The chemistry of these kinds of objects was barely known even for the case of the \rrp, which is by far the best-studied object. Apart from CO, \citet{bujarrabal2016} discovered \htcn\, \cuatrotres, \textsc{C\,i}\,\dosuno\ and \unocero, and \textsc{C\,ii}\,\dosuno.
In this work, we detected \cdsop\ and \cdoop\ \dosuno, and SO $J_{N}=6_{6}-5_{4}$, and we also tentatively detected HCN \unocero\ and \agua\ \tragua.
Both SO and \agua\ (at 22\,GHz) are good tracers of O-rich environments (see \sect\ref{corich}).
Additionally, we attained very significant SiO and SiS emission upper limits.
We estimated the upper limits on the abundance of these molecules following the method described in \app\ref{ap_abundancias}, assuming a 3$\sigma$ limit for intensity and integrated intensity lines and $T_{rot}=$ 11\mm 3\,K (see \tab\ref{lineas_rr} and \sect\ref{abundancias}). We find $\langle X \rangle _{SiO} <$ (5.9\mm 0.8)\x\xd{-9} and $\langle X \rangle _{SiS} <$ (2.3\mm 0.1)\x\xd{-10} (see \tab\ref{abun} for comparison).

The solar abundance of Si is $\sim$\,6.5\x\xd{-5} \citep[see e.g.,][]{asplund2021}. 
In the case of O-rich AGB stars, SiO abundances vary in between 4\x\xd{-5} and 5\x\xd{-6}, depending on the state of depletion onto dust grains \citep{verbena2019}.
These values for the abundances are considerably larger than the estimated upper limits for SiO and SiS.
Taking into account the fact that the abundance of SiO decreases as grains of dust are formed \citep{gobrecht2016}, it seems that the depletion process is very efficient, at least in the \rr and practically all Si could be depleted onto the grains. This conclusion is reinforced if we consider that the molecular emission of the \rr practically comes from the Keplerian disk, where depletion should be more effective than in the case of an expanding and very large envelope, as for AGB stars (since, in a rotating disk structure, there is more time for the grains to grow bigger).

\textsc{C\,ii} emission must have its origin in the PDR, because it is the best tracer of these regions. \textsc{C\,i} is also very often associated with PDRs.
The presence of \htcn\ is also compatible with the development of a PDR in the dense disk region that is close to the stellar system, which is a binary system with a secondary star and an accretion disk that emits in the UV \citep{witt2009,thomas2013}. The UV emission can excite a zone in between the \textsc{H\,ii} region and the Keplerian molecule-rich disk \citep{bujarrabal2016}.
According to models of the PDR chemistry, the origin of the HCN can be the photoinduced formation \citep{agundez2008} and it has been detected in the innermost regions of several disks around young stars. This fact is also consistent with the detection of HCN (see \fig\ref{fig:rr_mol}).

In conclusion, we think that the \rr most likely presents a O-rich gas chemistry, but it also presents a PDR in the innermost region of the Keplerian disk. The UV excess can explain the detection of HCN (and H$^{13}$CN, \textsc{C\,i}, and \textsc{C\,ii}) together with PAHs, even in an O-rich environment. We note that this is the typical situation in star-forming regions and that these are thought to be O-rich despite the presence of C-bearing species, including PAHs.

\begin{figure}[h]
\centering
%width=\sz\linewidth
\includegraphics[width=1 \linewidth]{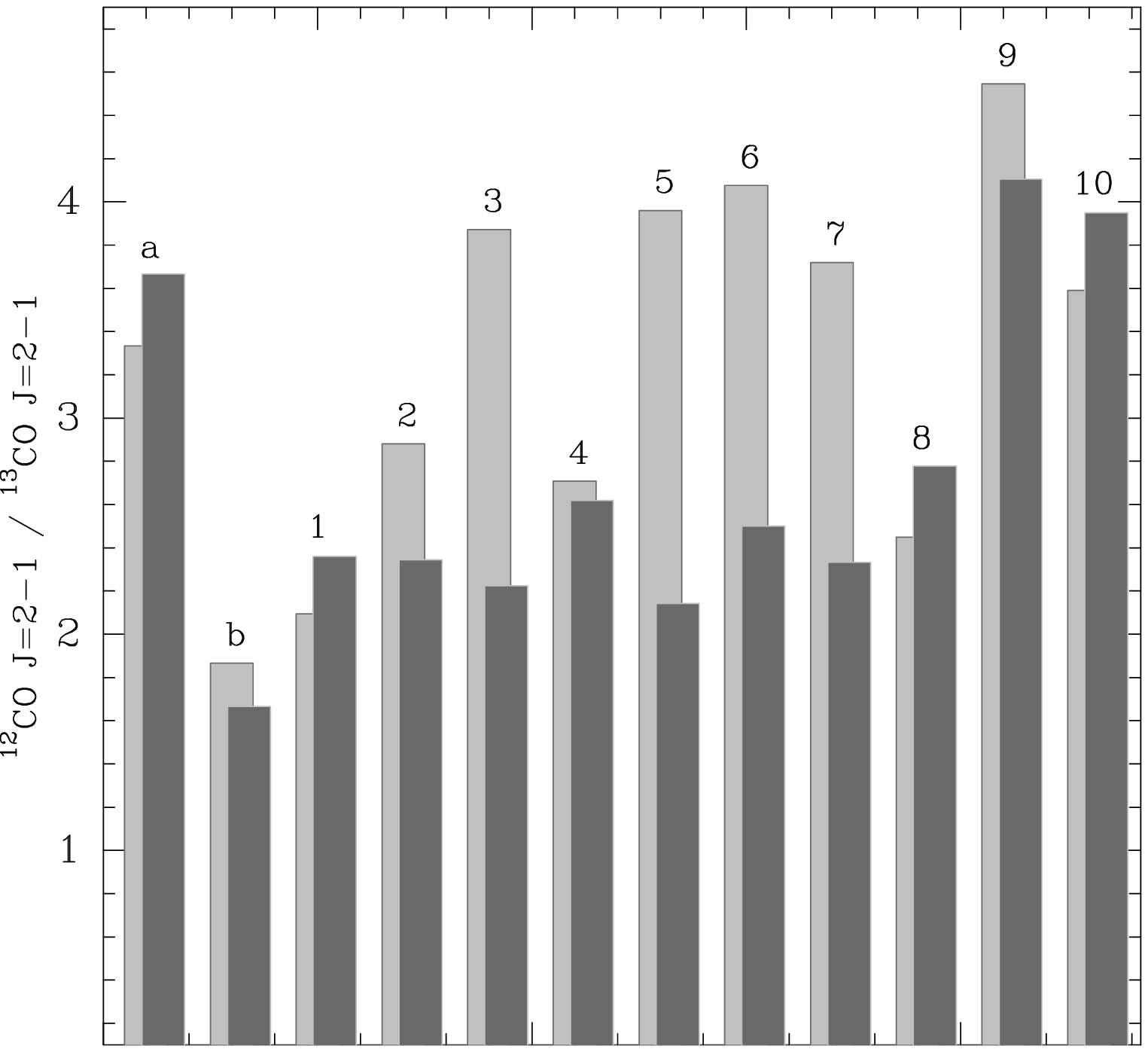} \\
\vspace{5mm}
\includegraphics[width=1\linewidth]{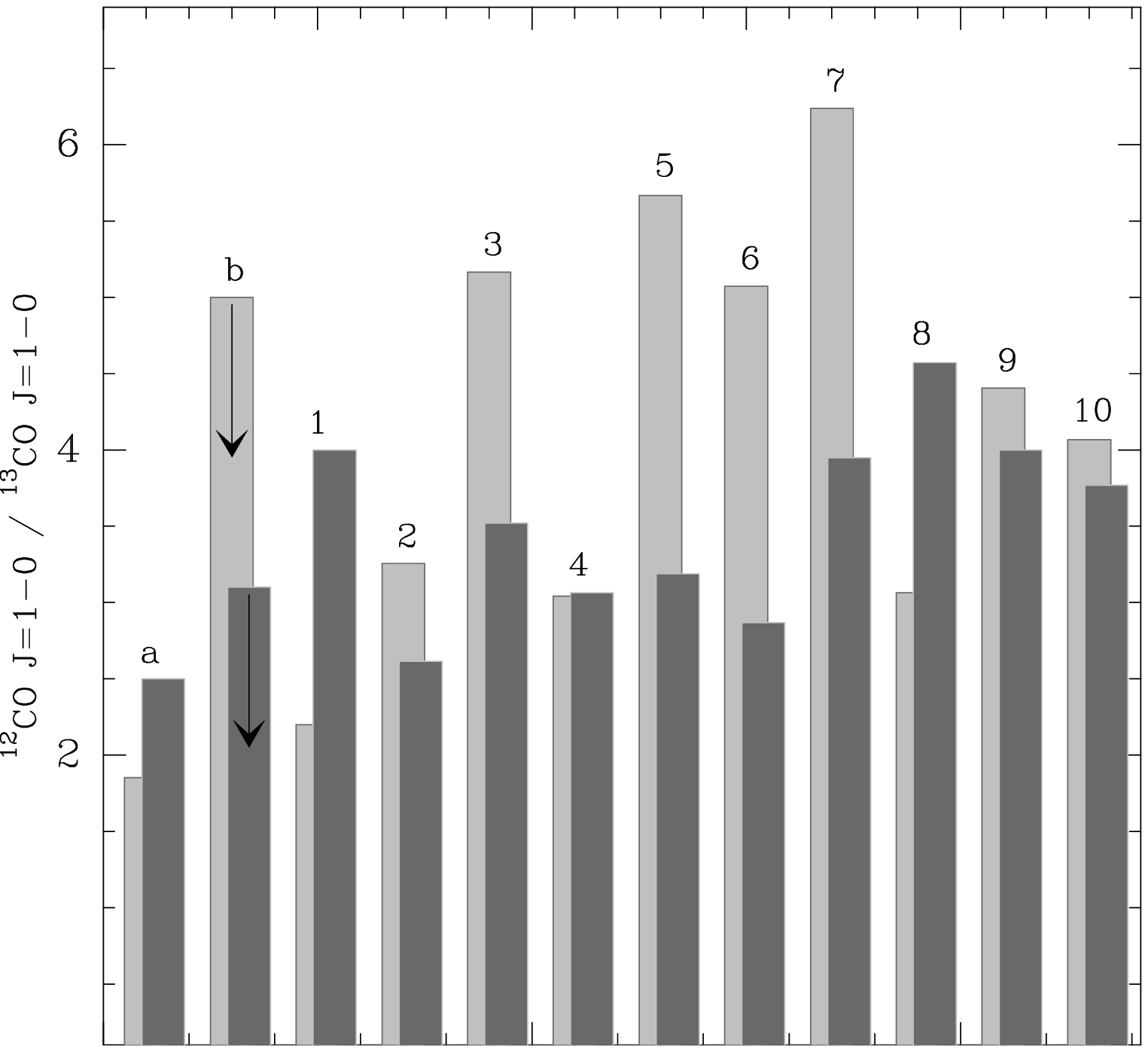}
\caption{\small Ratios of integrated intensities of \doce\,/\,\trece  in binary \pagb stars in light grey. We also show line-peak ratios in dark grey (\dosuno\ at \textit{top}; \unocero\ at \textit{bottom}).
Sources are ordered by outflow dominance and enumerated as follows: a\,$-$\,HR\,4049, b\,$-$\,DY\,Ori, 1\,$-$\,\acp, 2\,$-$\,\rrp, 3\,$-$\,\onp, 4\,$-$\,HD\,52961, 5\,$-$\,IRAS\,19157$-$0257, 6\,$-$\,IRAS\,18123+0511, 7\,$-$\,\irasp, 8\,$-$\,\aip, 9\,$-$\,IRAS\,20056+1834, and 10\,$-$\,\rsp. Sources a and b are not included in this mm-wave survey. Sources 1, 2, a, and b are disk-dominated binary \pagb stars, sources 6 to 10 are outflow-dominated, and sources 3, 4, and 5 are intermediate cases.}
    \label{fig:ratio_co}  
\end{figure}

\begin{table}[h]
\caption{O-isotopic abundance ratio, initial stellar mass estimation compared with the central total stellar mass and the total nebula mass.}
\small
%\tiny
%\centering
\vspace{-5mm}
\begin{center}
%\resizebox{\columnwidth}{!}{
\begin{tabular}{lcccc}

\hline \hline
\noalign{\smallskip}
%&&&&&\multicolumn{4}{|c|}{}\\
 %\\[-2ex]
\multirow{2}{*}{Source} & \multirow{2}{*}{$^{17}$O\,/\,$^{18}$O} & $M_{i}$ & $M_{Tc}$ & $M_{neb}$ \\
 & & [M$_{\odot}$] & [M$_{\odot}$] & [M$_{\odot}$] \\

\hline
\\[-2ex]
\vspace{1mm} 

\rr      & 0.8   &  1.5   &   1.7 & 1.4\x\xd{-2} \\             
%\vspace{1mm}               
89\,Herculis       & 0.3   &   1.2   &  1.7   & 1.4\x\xd{-2} \\
          
\hline
\end{tabular}
%}
% \\[1ex]

\end{center}
\small
\vspace{-1mm}
\textbf{Notes.}  $M_{i}$ represents the initial stellar mass of the \pagb component, $M_{Tc}$ is of the total central stellar mass, and $M_{neb}$ is the total mass of the nebula (Keplerian disk and outflow). Values of $M_{neb}$, for the \rr and \onp, are taken from \citet{bujarrabal2016,gallardocava2021}.
\label{masa_inicial}
\end{table}

\subsection{Isotopic ratios} \label{ratios}

\subsubsection{$^{17}$O\,/\,$^{18}$O}
Complex stellar evolution models reveal that the $^{17}$O\,/\,$^{18}$O ratio remains approximately constant over the entire thermally pulsating AGB phase, and the initial stellar mass ($M_{i}$)  can be directly obtained from this ratio \citep[][]{denutte2017}.
We can estimate the $^{17}$O\,/\,$^{18}$O  abundance ratio using the C$^{17}$O and C$^{18}$O\  \dosuno\ line intensities corrected for different beam widths and Einstein coefficients, because both molecules present almost identical radial depth dependence  \citep[][]{visser2009}, low optical depths, and very similar and easily predictable excitations conditions.

We applied this criterium to our entire sample. According to our results in \tabs\ref{lineas_rr} and \ref{lineas_89her}, we estimate the abundance ratio for the \rr and \onp, respectively.
Initial stellar masses for the \pagbs stars are shown in \tab\ref{masa_inicial}. We additionally show the total stellar mass ($M_{Tc}$) derived from the Keplerian dynamics and the total nebula mass ($M_{neb}$), which includes the mass of the rotating disk and outflow \citep[see][]{bujarrabal2016,gallardocava2021}.
We note that the derived values of $M_{i}$ are consistent with $M_{Tc}$, which must be higher because of the presence of a companion.
The $^{17}$O\,/\,$^{18}$O ratio for \on is typical of a O-rich object, while its chemistry is typical of C-rich environments. We note that abundance models are set up for single stars and not for binary stars where it is very likely that mass transfer happen after the results of the evolution of the individual stars.

\subsubsection{\doce\,/\,\trece}

For our binary \pagb stars, we find a \doce\,/\,\trece integrated-line ratio, with a mean value of 4.1 and a standard deviation of 1.4 for the transition \unocero\ (3.3\mm0.9 for \dosuno).
This ratio can be compared with intensity line ratios of other \pagb stars, in which we find 10.4\mm7.6 \citep[see][]{palla2000,balser2002,sanchezcontreras2012}.
Although these ratios must be corrected for different beam widths, Einstein coefficients, and opacity (especially for the \doce\ \dosuno\ lines), our objects seem to present higher \trece\ abundance compared with other \pagb stars.
As stated before, the \doce\,/\,\trece abundance ratio was previously derived for seven of our sources \citep[see][]{bujarrabal2016,bujarrabal2017,bujarrabal2018,gallardocava2021}. From these works, we are able to find a \doce\,/\,\trece mean value of 8.6\mm2.4. We can compare this ratio to that of AGB stars: 12.7\mm5.8, 28.0\mm16.2, and 31.3\mm28.2 for M-, S-, and C-type stars, respectively \citep[see][]{ramstedtolofsson2014}. Again, our objects present a higher \trece\ abundance compared with any sub-class of AGB stars, and are only similar to those of M-type stars presenting very low abundance ratios.

Our results support the suggestion by \citet{ramstedtolofsson2014} that the binary nature of the sources is the origin of these very low \doce\,/\,\trece ratios found in other post-AGB stars, because all our sources are bona-fide binary systems in which a strong interaction between the stars and the circumbinary disk is expected \citep{vanwinckel2003}.

\section{Conclusions}
\label{conclusiones}

We present a very deep survey of mm-wave lines in binary \pagb stars, which is the first systematic study of this kind.
We show our single-dish observational results in \sect\ref{resultados}, where detected lines together with very relevant upper limits are presented.

Our molecular study allows us to verify that our sources present lower molecular intensities of all species than those typical of AGB sources (see \sect\ref{pobreza_molecular}). This fact is very significant in those binary \pagb stars for which the Keplerian disk is the dominant component (i.e., the \rr and \acp) or at least it has a significant weight in the whole nebula, such as \onp. This lower molecular richness is also present in CO, and it is again very remarkable in disk-dominated sources. On the other hand, we find some overabundance of \trecep, especially in our outflow-dominated objects.

Our analysis of integrated intensities ratios of pairs of molecules other than CO leads us to classify the chemistry of (some of) our sources as C- or O-rich (see \sect\ref{corich}). We cataloged the nebulae around \acp, the \rrp, \aip, \rsp, and IRAS\,20056+1834 as O-rich environments. On the contrary, the nebula around \on is a C-rich environment.

We estimated abundances of the different species using the method described in detail in  \app\ref{ap_abundancias}. As expected, our results reveal that our sources present very low abundances in molecules other than CO (see \sect\ref{abundancias}) compared with standard evolved stars. 
The chemistry of these objects is yet to be studied from the theoretical point of view.

We also studied the isotopic ratios of our sources (see \sect\ref{ratios}). We deduced the initial stellar mass for the \rr and \on via the $^{17}$O\,/\,$^{18}$O ratio, which is consistent with the current central total stellar mass derived of our mm-wave interferometric maps and models (see \tab\ref{masa_inicial}). We studied the \doce\,/\,\trece ratios and we find that our sources, which are all \pagb stars, tend to present an overabundance of \trece compared to AGB and other \pagb stars.

\begin{acknowledgements}
We are grateful to the anonymous referee for the relevant recommendations and comments.
This work is based on observations carried out with the 30\,m\,IRAM telescope. IRAM is supported by INSU/CNRS (France), MPG (Germany), and IGN (Spain). 
This work is also based on observations with the 40\,m\, Yebes telescope of the National Geographic Institute of Spain (IGN) at Yebes Observatory. Yebes Observatory thanks the ERC for funding support under grant ERC-2013-Syg-610256-NANOCOSMOS.
This work is part of the AxiN and EVENTs\,/\,NEBULAE\,WEB research programs supported by Spanish AEI grants AYA\,2016-78994-P and PID2019-105203GB-C21, respectively.
IGC acknowledges Spanish MICIN the funding support of BES2017-080616.

\end{acknowledgements}

\bibliographystyle{aa}
\bibliography{referencias.bib}

%\newpage
\appendix

\section{Additional figures}

\begin{figure*}[h]
\center
\includegraphics[width=1 \linewidth]{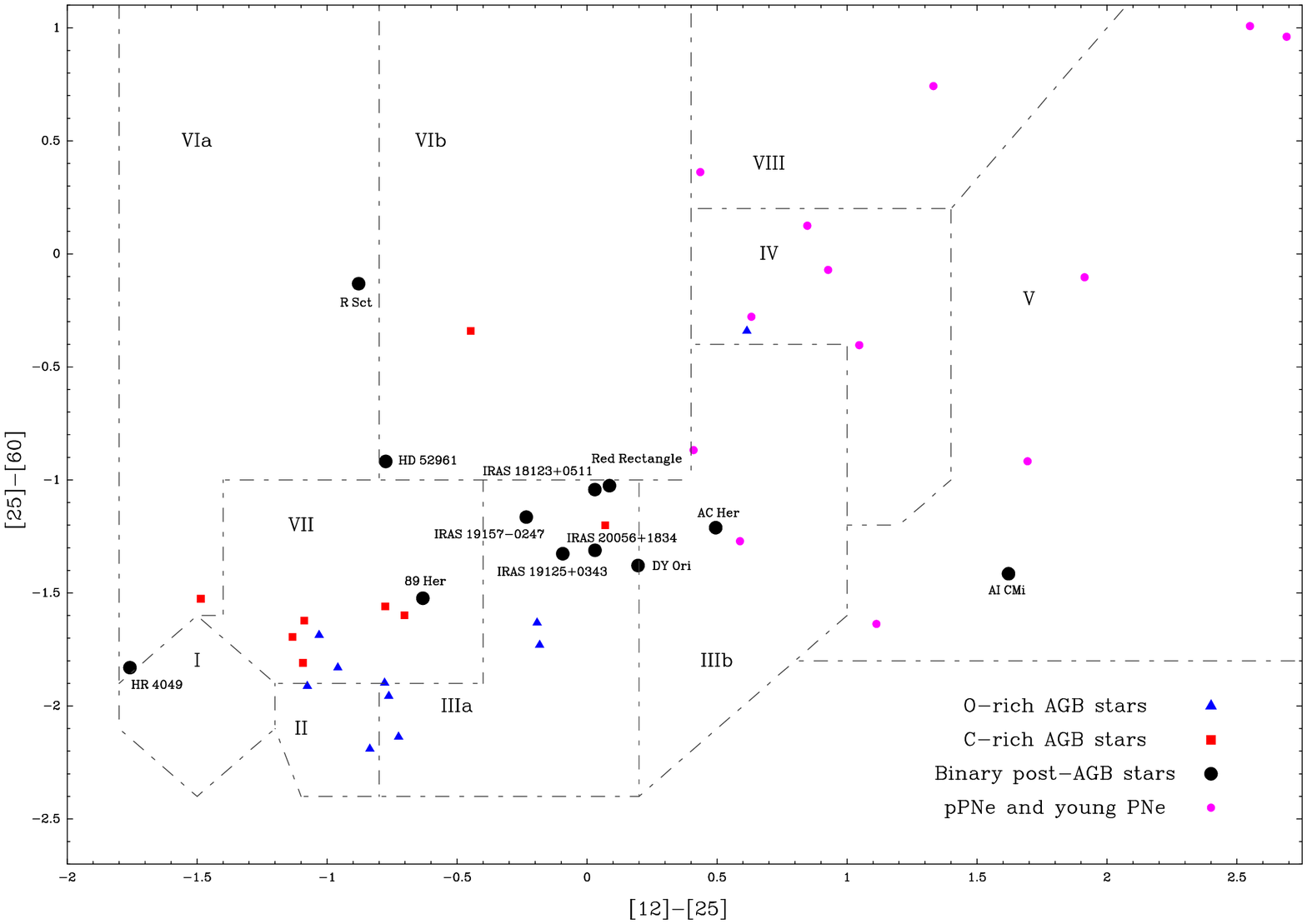}
\caption{\small 
Diagram of [25]$-$[60] vs. [12]$-$[25] IRAS colors. The IRAS color-color diagram is divided in different regions where sources with similar characteristics are located \citep[see][for details]{veenhabing1988}. Standard AGB stars are represented with blue triangles (O-rich) and red squares (C-rich), pre--planetary and young planetary nebulae are represented with purple circles and the binary post-AGB stars of this work are shown with black circles.}
    \label{fig:diagramacolorcolor}  
\end{figure*}

\begin{figure}[H]
\centering
%width=\sz\linewidth
\includegraphics[width=\sz\linewidth]{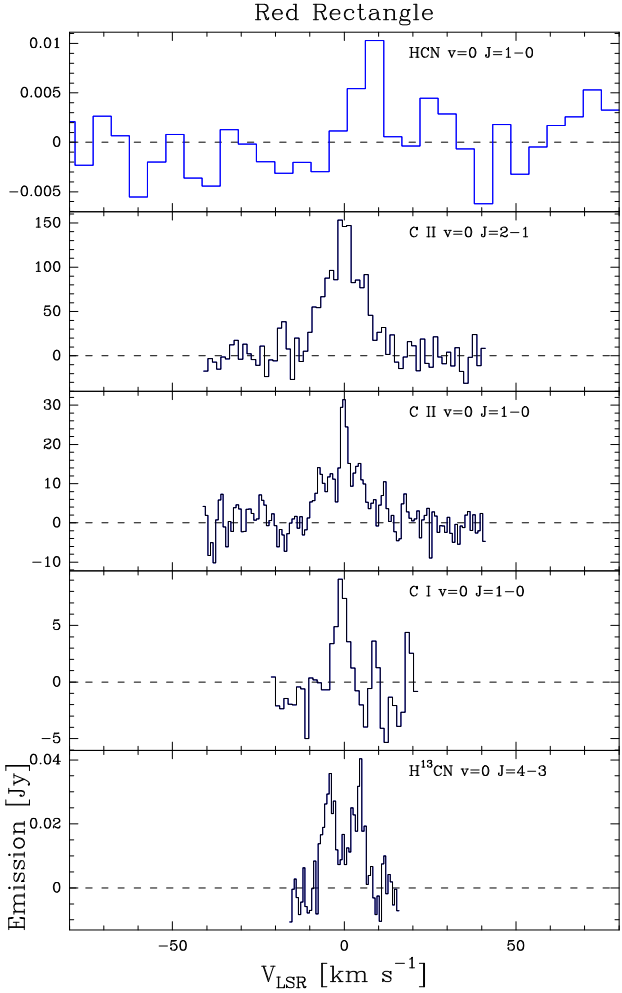}
\caption{\small Spectra for HCN \unocero\ detected in the \rrp. We also show other PDR-tracing lines in black taken from \citet{bujarrabal2016}. The x-axis indicates velocity with respect to the local standard of rest ($V_{LSR}$) and the y-axis represents the emitted emission measured in Jansky.}
    \label{fig:rr_mol_carb}  
\end{figure}

In this Appendix, we show other relevant figures.
In \fig\ref{fig:rr_mol_carb} we present our tentative detection of HCN \unocero\ in the \rrp. We also show in black H$^{13}$CN \unocero, \textsc{C\,i} \unocero, \textsc{C\,ii} \unocero, and \textsc{C\,ii} \dosuno\ \citep[ALMA data taken from][]{bujarrabal2016}.

In \fig\ref{fig:diagramacolorcolor}, we show the color-color diagram where we represent [25]$-$[60] versus [12]$-$[25] IRAS colors. We also show the regions populated stars with similar properties \citep[see][for more details]{veenhabing1988}. These kinds of plots show the evolutionary sequence of AGB\,/\,\pagb stars. We note that the sources of this work, the binary post-AGB stars, are in between of AGB stars and pPNe and young PNe, which confirms the nature of our sources.

\section{Estimation of abundances: Methodology} \label{ap_abundancias}

In this appendix, we show in detail the procedure used for the estimation of the abundance of molecules with more than one detected transition, assuming optically thin emission. In our case, we use this procedure to estimate the abundance of SiO for \rs and \ai (see \sects\ref{resultados_aicmi} and \ref{resultados_rsct}).

The observed main-beam temperature for a specific molecular transition is expressed as:

\begin{equation} \label{eq:tmb}
T_{mb}\,(V_{LSR})=S_{\nu} \,\tau \, \phi\,(V_{LSR}) \, \frac{\Omega_{s}}{\Omega_{mb}},
\end{equation}
where $\tau$ is the characteristic optical depth, $\phi\, (V_{LSR})$ is the normalized line profile in terms of Doppler-shifted velocity, and $\Omega_{s}$ and $\Omega_{mb}$ are the solid angles subtended by the source and the telescope main beam, respectively. These are given by:
\begin{equation}
\Omega_{mb} = \frac{\pi}{4\ln{2}} \, \theta_{HPBW}^{2} \cong 1.133\,\theta_{HPBW}^{2},
\end{equation}
\vspace{-5mm}
\begin{equation}
\Omega_{s} = 2\pi\, (1-\cos \alpha) \cong \pi \alpha^{2} \cong \pi \frac{r^{2}}{D^{2}}= \frac{S}{D^{2}} ,
\end{equation}
where $\theta_{HPBW}$ is the Half Power Beam Width (HPBW) and $\alpha$ is the angle formed by the radius $r$ of the source from the observer at a distance $D$. In the limit of small angle $\alpha$, we assume that $\tan \alpha  \cong \frac{r}{D}$ and $\cos \alpha \cong 1 - \frac{\alpha^{2}}{2} $.
The ratio $\frac{\Omega_{s}}{\Omega_{mb}}$ is known as the dilution factor, which can only take values lower than 1 (for $\Omega_{s}$ larger than $\Omega_{mb}$, namely, spatially resolved sources, the dilution factor is set to one).

The source function is given by:
\begin{equation}
S_{\nu}=\frac{h\nu}{k_{B}} \left( \frac{1}{e^{\frac{h\nu}{k_{B} T_{ex}}}-1} -  \frac{1}{e^{\frac{h\nu}{k_{B} T_{BG}}}-1}  \right) , 
\end{equation}
where $T_{ex}$ is the excitation temperature of the transition and $T_{BG}=2.7$\,K is the background temperature. The optical depth, $\tau$, can be expressed as:
\begin{equation} \label{eq:tau}
\tau=\frac{c^{3}}{8\pi\nu^{3}} \, A_{ul} \, g_{u} \, (x_{l}-x_{u}) \, n_{T} \, X \, l ,
\end{equation}
where $X$ is the abundance of the studied molecule, $n_{T}$ is the total number density of particles, and $l$ is the typical length of the source along the line of sight. The subindexes $u$ and $l$ represent the upper and lower levels of the transition, respectively. $A_{ul}$ is the Einstein coefficient of the transition, $g_{u}$ is the statistical weight of the upper limit expressed as
\begin{equation}
g_{J} = 2\,J+1,
\end{equation}
and $x_{u}$ and $x_{l}$ are:
\begin{equation}
x_{J} \sim \frac{e^{-\frac{E_{J}}{T_{rot}}}}{F\,(T_{rot})}  , 
\end{equation}
where $E$ is the energy of the considered levels, in terms of temperature, and for simple linear rotors it can be calculated as:
\begin{equation}
E_{J}=B_{rot}\, J\,(J+1),
\end{equation}
where $B_{rot}$ is the rotational constant of the molecule. $T_{rot}$ is the rotational temperature and it 
is  defined  as  the  typical  value  or  average  of  the excitation  temperatures  of  the  relevant  rotational  lines  (in  our case, low-$J$ transitions) and is used to approximately calculate the partition function $F\,(T_{rot})$:
\begin{equation}
F\,(T_{rot}) = \sum_{J} g_{J}\,e^{-\frac{E_{J}}{T_{rot}}} = \frac{T_{rot}}{B_{rot}}.
\end{equation}
We can express $\tau$ (\eq\ref{eq:tau}) in terms of the mass, taking into account:
\begin{equation}
M = n_{T}\,l\,S = n\,M_{mol}\,l\,S
\end{equation}

Integrating (\eq\ref{eq:tmb}) in terms of velocity, we get:

\begin{equation*}
\int T_{mb}\,(V_{LSR})\,dV = 
\end{equation*}
\vspace{-5mm}
\begin{equation} \label{eq:tmb_masa}
 S_{\nu}  \frac{c^{3}}{8\pi\nu^{3}} A_{ul} \, g_{u} \, (x_{l}-x_{u}) \, X \frac{4\ln{2}}{\pi} \, \theta_{HPBW}^{-2} \, M_{mol} \, D^{-2}
\end{equation}

We estimate $T_{rot}$ from the line intensity ratio for those molecules, where we  observed more than one transition (SiO, see \sect\ref{resultados}). We calculated several abundances for the different transitions ($X_{J=u-l}$, $X'_{J=u'-l'}$) of the same molecule and then we estimated the parameter $d_{i}$ for each pair of transitions:

\begin{equation} \label{eq:cantidad_d}
d_{i}\,(T_{rot}) = \frac{\left | X_{J=u-l} - X'_{J=u'-l'} \right |}{(X_{J=u-l} + X'_{J=u'-l'})/2}.
\end{equation}

We can quantify the quality of the fit through the parameter $\Delta d\,(T_{rot})$:

\begin{equation} \label{eq:error}
\Delta d\,(T_{rot}) = \left( \prod_{i}^{N} d_{i}\,(T_{rot}) \right)^{\frac{1}{N}} , 
\end{equation}
where $N$ indicates the number of transitions used in each case.
The value of $T_{rot}$ that minimizes the value of $\Delta d\,(T_{rot})$ in \eq\ref{eq:error} will be the best-fit rotational temperature (see \fig\ref{fig:abundancia_media}).
Finally, knowing $T_{rot}$, we can estimate the abundance through \eq\ref{eq:tmb_masa}.

Additionally, using  \eq\ref{eq:tmb}, we can express the optical depth as:
\begin{equation} \label{eq:tau_simp}
\tau \sim \frac{T_{mb}}{T_{ex}} \frac{\Omega_{mb}}{\Omega_{s}}.
\end{equation}
and we can express the averaged optical depth over the beam of the telescope as:
\begin{equation} \label{eq:tau_simp_simp}
\overline{\tau} \sim \frac{T_{mb}}{T_{ex}}.
\end{equation}

The averaged optical depths values can be seen in \tab\ref{tau_calc}. We note that these $\overline{\tau}$ values are very small, and that for the dilution factors expected for our sources, always lower than 10 \citep[see][]{gallardocava2021,bujarrabal2016,bujarrabal2013a}, implying that the lines are optically thin ($\tau< 1$).
Additionally, we note that always compare our detections with \trece\dosuno, which is an optically thin line and shows a similar beam size.

\begin{table}[h]
\caption{\textbf{Optical depths estimated.}}
\small
%\tiny
%\centering
\vspace{-5mm}
\begin{center}
%\resizebox{\columnwidth}{!}{
\begin{tabular}{llcc}

\hline \hline
\noalign{\smallskip}
%&&&&&\multicolumn{4}{|c|}{}\\
 %\\[-2ex]
\multirow{2}{*}{Source} & \multirow{2}{*}{Molecule} & $T_{rot}$  &  \multirow{2}{*}{$\overline{\tau}$} \\
 & & [K] &  \\

\hline
\\[-2ex]
%\vspace{1mm} 

AI\,CMi       & SiO          &  8                   &  7.3\x\xd{-3}  \\
              & $^{29}$SiO   &  8                   &  6.5\x\xd{-4}  \\
\vspace{1mm} 
              & $^{30}$SiO   &  8               &  <\,3.0\x\xd{-4}  \\

R\,Scuti      & SiO          &  14                  &  1.9\x\xd{-2}  \\
              & $^{29}$SiO   &  14                 &  1.1\x\xd{-3}  \\
              & $^{30}$SiO   &  14                 &  8.7\x\xd{-4}  \\
\vspace{1mm}               
              & HCO$^{+}$    &  14               & 6.0\x\xd{-4}   \\

 \rr           & HCN          &  8\,$-$\,14  &  (2.6\mm 0.7)\x\xd{-4} \\
%\vspace{1mm} 
89\,Herculis  & HCN          &  8\,$-$\,14    & (2.4\mm 0.6)\x\xd{-4} \\
              & SiS          &  8\,$-$\,14    & (2.9\mm 0.8)\x\xd{-4} \\

              & CS           &  8\,$-$\,14    & (9.6\mm 2.6)\x\xd{-4} \\

\hline
\end{tabular}
%}
% \\[1ex]

\end{center}
\small
\vspace{-1mm}
\textbf{Notes.} The averaged optical depth, $\overline{\tau}$, is the value of $\tau$ averaged over the telescope beam, see \eqs\ref{eq:tau_simp} and \ref{eq:tau_simp_simp}.
\label{tau_calc}
\end{table}

\section{Tables} \label{tablas_completas}

In this appendix, we show the most relevant line parameters for \allsp (see \sect\ref{resultados}). We note that most of the sources have been observed in the 1.3, 2, 3, 7, and 13\,mm bands (see \tab\ref{rangos}).

\begin{table*}[h]
\caption{Radio molecular line survey of \acp.}
\small
\tiny
%\centering
\vspace{-5mm}
\begin{center}

\begin{tabular*}{\textwidth}{@{\extracolsep{\fill\quad}}ll @{\hspace{0.5ex}} lccccccl}
\hline \hline
\noalign{\smallskip}
%&&&&&\multicolumn{4}{|c|}{}\\
 %\\[-2ex]

 \multirow{2}{*}{Molecule} & \multicolumn{2}{c}{\multirow{2}{*}{Transition}} & I\,(peak) & $\sigma$ & $\int I\,dV$ & $\sigma \left(\int I\,dV \right)$  & Sp.\,Res. & $V_{LSR}$ &  \multirow{2}{*}{Comments} \\
 &  & & [Jy] & [Jy] & [Jy\,km\,s$^{-1}$] & [Jy\,km\,s$^{-1}$] & [km\,s$^{-1}$]  & [km\,s$^{-1}$] & \\
\hline
\\[-2ex]

       \cdoop         &              \vcero         &             \dosuno         &   $\leq$\,4.6E-02         &   1.5E-02         &   $\leq$\,1.0E-01         &   3.4E-02         &      0.53         &   $-$         &   Not detected        \\
              \cdsop         &              \vcero         &             \dosuno         &   $\leq$\,4.6E-02         &   1.5E-02         &   $\leq$\,1.0E-01         &   3.4E-02         &      0.52         &   $-$         &   Not detected        \\
                 SiO         &              \vcero         &            \unocero         &   $\leq$\,2.2E-02         &   7.3E-03         &   $\leq$\,4.9E-02         &   1.6E-02         &      0.53         &   $-$         &   Not detected        \\
                             &                             &             \dosuno         &   $\leq$\,1.6E-02         &   5.2E-03         &   $\leq$\,3.5E-02         &   1.2E-02         &      0.67         &   $-$         &   Not detected        \\
                             &                             &        \cincocuatro         &   $\leq$\,5.4E-02         &   1.8E-02         &   $\leq$\,1.2E-01         &   4.0E-02         &      0.54         &   $-$         &   Not detected        \\
\hline
\end{tabular*}
% \\[1ex]

\end{center}
\small
%\vspace{-1mm}
%\textbf{Notes.} This table collects the main molecular transitions for \acp.

\label{lineas_acher}
\end{table*}

\begin{table*}[h]
\caption{Radio molecular line survey of the Red\,Rectangle.}
\small
\tiny
%\centering
\vspace{-5mm}
\begin{center}

\begin{tabular*}{\textwidth}{@{\extracolsep{\fill\quad}}ll @{\hspace{0.5ex}} lccccccl}
\hline \hline
\noalign{\smallskip}
%&&&&&\multicolumn{4}{|c|}{}\\
 %\\[-2ex]

 \multirow{2}{*}{Molecule} & \multicolumn{2}{c}{\multirow{2}{*}{Transition}} & I\,(peak) & $\sigma$ & $\int I\,dV$ & $\sigma \left(\int I\,dV \right)$  & Sp.\,Res. & $V_{LSR}$ &  \multirow{2}{*}{Comments} \\
 &  & & [Jy] & [Jy] & [Jy\,km\,s$^{-1}$] & [Jy\,km\,s$^{-1}$] & [km\,s$^{-1}$]  & [km\,s$^{-1}$] & \\
\hline
\\[-2ex]

              \cdoop         &              \vcero         &             \dosuno         &   1.7E-01         &   2.4E-02         &   9.5E-01         &   6.5E-02         &      0.53         &     $-$0.64         &                       \\
              \cdsop         &              \vcero         &             \dosuno         &   1.8E-01         &   2.4E-02         &   7.9E-01         &   4.8E-02         &      0.52         &     $-$0.59         &                       \\
                 SiO         &              \vcero         &            \unocero         &   $\leq$\,2.4E-02         &   8.0E-03         &   $\leq$\,5.3E-02         &   1.8E-02         &      0.53         &   $-$         &    Not detected        \\
                             &                             &             \dosuno         &   $\leq$\,2.2E-02         &   7.5E-03         &   $\leq$\,4.9E-02         &   1.6E-02         &      0.67         &   $-$         &   Not detected        \\
                    & & \cincocuatro & $-$ & $-$ & $-$ & $-$ & $-$ & $-$ & Not observed \\
\hline
\end{tabular*}
% \\[1ex]

\end{center}
\small
%\vspace{-1mm}
%\textbf{Notes.} Same as in \tab\ref{lineas_acher} but for the \rrp.

\label{lineas_rr}
\end{table*}

\begin{table*}[h]
\caption{Radio molecular line survey of \onp.}
\small
\tiny
%\centering
\vspace{-5mm}
\begin{center}

\begin{tabular*}{\textwidth}{@{\extracolsep{\fill\quad}}ll @{\hspace{0.5ex}} lccccccl}
\hline \hline
\noalign{\smallskip}
%&&&&&\multicolumn{4}{|c|}{}\\
 %\\[-2ex]

 \multirow{2}{*}{Molecule} & \multicolumn{2}{c}{\multirow{2}{*}{Transition}} & I\,(peak) & $\sigma$ & $\int I\,dV$ & $\sigma \left(\int I\,dV \right)$  & Sp.\,Res. & $V_{LSR}$ &  \multirow{2}{*}{Comments} \\
 &  & & [Jy] & [Jy] & [Jy\,km\,s$^{-1}$] & [Jy\,km\,s$^{-1}$] & [km\,s$^{-1}$]  & [km\,s$^{-1}$] & \\
\hline
\\[-2ex]

              \cdoop         &              \vcero         &             \dosuno         &   1.2E-01         &   1.8E-02         &   2.0E-01         &   2.0E-02         &      0.27         &     -8.36         &                       \\
               \cdsop         &              \vcero         &             \dosuno         &   1.3E-02         &   5.1E-03         &   6.1E-02         &   5.2E-02         &      4.17         &    -11.10         &      Tentative        \\
                 SiO         &              \vcero         &            \unocero         &   $\leq$\,2.3E-02         &   7.8E-03         &   $\leq$\,4.3E-02         &   1.4E-02         &      0.53         &   $-$         &   Not detected        \\
                             &                             &             \dosuno         &   $\leq$\,1.7E-02         &   5.8E-03         &   $\leq$\,3.2E-02         &   1.1E-02         &      0.67         &   $-$         &   Not detected        \\
                             &                             &        \cincocuatro         &   $\leq$\,9.4E-02         &   3.1E-02         &   $\leq$\,1.7E-01         &   5.8E-02         &      0.54         &   $-$         &   Not detected        \\
\hline
\end{tabular*}
% \\[1ex]

\end{center}
\small
%\vspace{-1mm}
%\textbf{Notes.} Same as in \tab\ref{lineas_acher} but for \onp.

\label{lineas_89her}
\end{table*}

\begin{table*}[h]
\caption{Radio molecular line survey of HD\,52961.}
\small
\tiny
%\centering
\vspace{-5mm}
\begin{center}

\begin{tabular*}{\textwidth}{@{\extracolsep{\fill\quad}}ll @{\hspace{0.5ex}} lccccccl}
\hline \hline
\noalign{\smallskip}
%&&&&&\multicolumn{4}{|c|}{}\\
 %\\[-2ex]

 \multirow{2}{*}{Molecule} & \multicolumn{2}{c}{\multirow{2}{*}{Transition}} & I\,(peak) & $\sigma$ & $\int I\,dV$ & $\sigma \left(\int I\,dV \right)$  & Sp.\,Res. & $V_{LSR}$ &  \multirow{2}{*}{Comments} \\
 &  & & [Jy] & [Jy] & [Jy\,km\,s$^{-1}$] & [Jy\,km\,s$^{-1}$] & [km\,s$^{-1}$]  & [km\,s$^{-1}$] & \\
\hline
\\[-2ex]

              \cdoop         &              \vcero         &             \dosuno         &   $\leq$\,1.3E-01         &   4.3E-02         &   $\leq$\,3.3E-01         &   1.1E-01         &      0.53         &   $-$         &   Not detected        \\
              \cdsop         &              \vcero         &             \dosuno         &   $\leq$\,1.3E-01         &   4.4E-02         &   $\leq$\,3.3E-01         &   1.1E-01         &      0.52         &     $-$        &      Not detected        \\
                 SiO         &              \vcero         &            \unocero         &   $\leq$\,1.8E-02         &   6.0E-03         &   $\leq$\,4.5E-02         &   1.5E-02         &      0.53         &   $-$         &   Not detected        \\
                             &                             &             \dosuno         &   $\leq$\,7.7E-02         &   2.6E-02         &   $\leq$\,1.9E-01         &   6.4E-02         &      0.67         &   $-$         &   Not detected        \\
                             &                             &        \cincocuatro         &   $\leq$\,1.5E-01         &   4.9E-02         &   $\leq$\,3.7E-01         &   1.2E-01         &      0.54         &   $-$         &   Not detected        \\
\hline
\end{tabular*}
% \\[1ex]

\end{center}
\small
%\vspace{-1mm}
%\textbf{Notes.} Same as in \tab\ref{lineas_acher} but for HD\,52961.

\label{lineas_hd52961}
\end{table*}

\begin{table*}[h]
\caption{Radio molecular line survey of IRAS\,19157$-$0257.}
\small
\tiny
%\centering
\vspace{-5mm}
\begin{center}

\begin{tabular*}{\textwidth}{@{\extracolsep{\fill\quad}}ll @{\hspace{0.5ex}} lccccccl}
\hline \hline
\noalign{\smallskip}
%&&&&&\multicolumn{4}{|c|}{}\\
 %\\[-2ex]

 \multirow{2}{*}{Molecule} & \multicolumn{2}{c}{\multirow{2}{*}{Transition}} & I\,(peak) & $\sigma$ & $\int I\,dV$ & $\sigma \left(\int I\,dV \right)$  & Sp.\,Res. & $V_{LSR}$ &  \multirow{2}{*}{Comments} \\
 &  & & [Jy] & [Jy] & [Jy\,km\,s$^{-1}$] & [Jy\,km\,s$^{-1}$] & [km\,s$^{-1}$]  & [km\,s$^{-1}$] & \\
\hline
\\[-2ex]

                 SiO         &              \vcero         &            \unocero         &   $\leq$\,4.2E-02         &   1.4E-02         &   $\leq$\,1.3E-01         &   4.4E-02         &      0.53         &   $-$         &   Not detected        \\
                             &                             &             \dosuno         &   $\leq$\,1.5E-01         &   5.1E-02         &   $\leq$\,4.7E-01         &   1.6E-01         &      0.67         &   $-$         &   Not detected        \\
                             &               \vuno         &            \unocero         &   $\leq$\,4.1E-02         &   1.4E-02         &   $\leq$\,1.3E-01         &   4.2E-02         &      0.53         &   $-$         &   Not detected        \\
                             &                             &             \dosuno         &   $\leq$\,1.5E-01         &   5.0E-02         &   $\leq$\,4.6E-01         &   1.5E-01         &      0.68         &   $-$         &   Not detected        \\
                             &               \vdos         &            \unocero         &   $\leq$\,3.9E-02         &   1.3E-02         &   $\leq$\,7.2E-02         &   2.4E-02         &      0.53         &     $-$         &      Not detected        \\
\hline
\end{tabular*}
% \\[1ex]

\end{center}
\small
%\vspace{-1mm}
%\textbf{Notes.} Same as in \tab\ref{lineas_acher} but for IRAS\,19157$-$0257.

\label{lineas_iras19157}
\end{table*}

\begin{table*}[h]
\caption{Radio molecular line survey of IRAS\,18123+0511.}
\small
\tiny
%\centering
\vspace{-5mm}
\begin{center}

\begin{tabular*}{\textwidth}{@{\extracolsep{\fill\quad}}ll @{\hspace{0.5ex}} lccccccl}
\hline \hline
\noalign{\smallskip}
%&&&&&\multicolumn{4}{|c|}{}\\
 %\\[-2ex]

 \multirow{2}{*}{Molecule} & \multicolumn{2}{c}{\multirow{2}{*}{Transition}} & I\,(peak) & $\sigma$ & $\int I\,dV$ & $\sigma \left(\int I\,dV \right)$  & Sp.\,Res. & $V_{LSR}$ &  \multirow{2}{*}{Comments} \\
 &  & & [Jy] & [Jy] & [Jy\,km\,s$^{-1}$] & [Jy\,km\,s$^{-1}$] & [km\,s$^{-1}$]  & [km\,s$^{-1}$] & \\
\hline
\\[-2ex]

              \cdoop         &              \vcero         &             \dosuno         &   $\leq$\,2.6E-01         &   8.7E-02         &   $\leq$\,1.6E+00         &   5.4E-01         &      0.53         &   $-$         &   Not detected        \\
              \cdsop         &              \vcero         &             \dosuno         &   $\leq$\,2.8E-01         &   9.3E-02         &   $\leq$\,1.7E+00         &   5.8E-01         &      0.52         &   $-$         &   Not detected        \\
                 SiO         &              \vcero         &            \unocero         &   $\leq$\,5.0E-02         &   1.7E-02         &   $\leq$\,3.1E-01         &   1.0E-01         &      0.53         &   $-$         &   Not detected        \\
                             &                             &             \dosuno         &   $\leq$\,1.0E-01         &   3.3E-02         &   $\leq$\,6.2E-01         &   2.1E-01         &      0.67         &   $-$         &   Not detected        \\
                             &                             &        \cincocuatro         &   $\leq$\,3.2E-01         &   1.1E-01         &   $\leq$\,2.0E+00         &   6.5E-01         &      0.54         &   $-$         &   Not detected        \\
\hline
\end{tabular*}
% \\[1ex]

\end{center}
\small
%\vspace{-1mm}
%\textbf{Notes.} Same as in \tab\ref{lineas_acher} but for IRAS\,18123+0511.

\label{lineas_iras18123}
\end{table*}

\begin{table*}[h]
\caption{Radio molecular line survey of \irasp.}
\small
\tiny
%\centering
\vspace{-5mm}
\begin{center}

\begin{tabular*}{\textwidth}{@{\extracolsep{\fill\quad}}ll @{\hspace{0.5ex}} lccccccl}
\hline \hline
\noalign{\smallskip}
%&&&&&\multicolumn{4}{|c|}{}\\
 %\\[-2ex]

 \multirow{2}{*}{Molecule} & \multicolumn{2}{c}{\multirow{2}{*}{Transition}} & I\,(peak) & $\sigma$ & $\int I\,dV$ & $\sigma \left(\int I\,dV \right)$  & Sp.\,Res. & $V_{LSR}$ &  \multirow{2}{*}{Comments} \\
 &  & & [Jy] & [Jy] & [Jy\,km\,s$^{-1}$] & [Jy\,km\,s$^{-1}$] & [km\,s$^{-1}$]  & [km\,s$^{-1}$] & \\
\hline
\\[-2ex]
              \cdoop         &              \vcero         &             \dosuno         &   $\leq$\,1.7E-01         &   5.8E-02         &   $\leq$\,6.0E-01         &   2.0E-01         &      0.53         &   $-$         &   Not detected        \\
              \cdsop         &              \vcero         &             \dosuno         &   $\leq$\,1.5E-01         &   4.9E-02         &   $\leq$\,5.1E-01         &   1.7E-01         &      0.52         &   $-$         &   Not detected        \\
                 SiO         &              \vcero         &            \unocero         &   $\leq$\,2.1E-02         &   7.2E-03         &   $\leq$\,7.4E-02         &   2.5E-02         &      0.53         &   $-$         &   Not detected        \\
              &                             &             \dosuno                        &   $\leq$\,5.4E-02         &   1.8E-02         &   $\leq$\,1.8E-01         &   6.1E-02         &      0.67         &     $-$         &      Not detected        \\
                             &                             &        \cincocuatro         &   $\leq$\,2.2E-01         &   7.2E-02         &   $\leq$\,7.5E-01         &   2.5E-01         &      0.54         &   $-$         &   Not detected        \\
\hline
\end{tabular*}
% \\[1ex]

\end{center}
\small
%\vspace{-1mm}
%\textbf{Notes.} Same as in \tab\ref{lineas_acher} but for \irasp.

\label{lineas_iras19125}
\end{table*}

\begin{table*}[h]
\caption{Radio molecular line survey of \aip.}
\small
\tiny
%\centering
\vspace{-5mm}
\begin{center}

\begin{tabular*}{\textwidth}{@{\extracolsep{\fill\quad}}ll @{\hspace{0.5ex}} lccccccl}
\hline \hline
\noalign{\smallskip}
%&&&&&\multicolumn{4}{|c|}{}\\
 %\\[-2ex]

 \multirow{2}{*}{Molecule} & \multicolumn{2}{c}{\multirow{2}{*}{Transition}} & I\,(peak) & $\sigma$ & $\int I\,dV$ & $\sigma \left(\int I\,dV \right)$  & Sp.\,Res. & $V_{LSR}$ &  \multirow{2}{*}{Comments} \\
 &  & & [Jy] & [Jy] & [Jy\,km\,s$^{-1}$] & [Jy\,km\,s$^{-1}$] & [km\,s$^{-1}$]  & [km\,s$^{-1}$] & \\
\hline
\\[-2ex]

              \cdoop         &              \vcero         &             \dosuno         &   $\leq$\,1.0E-01         &   3.5E-02         &   $\leq$\,1.9E-01         &   6.4E-02         &      0.53         &     $-$         &      Not detected        \\
              \cdsop         &              \vcero         &             \dosuno         &   $\leq$\,9.5E-02         &   3.2E-02         &   $\leq$\,1.7E-01         &   5.8E-02         &      0.52         &   $-$         &   Not detected        \\
                 SiO         &              \vcero         &            \unocero         &   2.0E-02         &   5.4E-03         &   7.5E-02         &   2.0E-02         &      1.05         &     29.11         &                       \\
                             &                             &             \dosuno         &   1.1E-01         &   1.8E-02         &   8.8E-01         &   7.0E-02         &      0.67         &     29.02         &                       \\
                             &                             &        \cincocuatro         &   2.9E-01         &   3.7E-02         &   2.0E+00         &   1.0E-01         &      0.54         &     29.00         &                       \\
\hline
\end{tabular*}
% \\[1ex]

\end{center}
\small
%\vspace{-1mm}
%\textbf{Notes.} Same as in \tab\ref{lineas_acher} but for \aip.

\label{lineas_aicmi}
\end{table*}

\begin{table*}[h]
\caption{Radio molecular line survey of IRAS\,20056+1834.}
\small
\tiny
%\centering
\vspace{-5mm}
\begin{center}

\begin{tabular*}{\textwidth}{@{\extracolsep{\fill\quad}}ll @{\hspace{0.5ex}} lccccccl}
\hline \hline
\noalign{\smallskip}
%&&&&&\multicolumn{4}{|c|}{}\\
 %\\[-2ex]

 \multirow{2}{*}{Molecule} & \multicolumn{2}{c}{\multirow{2}{*}{Transition}} & I\,(peak) & $\sigma$ & $\int I\,dV$ & $\sigma \left(\int I\,dV \right)$  & Sp.\,Res. & $V_{LSR}$ &  \multirow{2}{*}{Comments} \\
 &  & & [Jy] & [Jy] & [Jy\,km\,s$^{-1}$] & [Jy\,km\,s$^{-1}$] & [km\,s$^{-1}$]  & [km\,s$^{-1}$] & \\
\hline
\\[-2ex]

              \cdoop         &              \vcero         &             \dosuno         &   $\leq$\,1.4E-01         &   4.7E-02         &   $\leq$\,4.6E-01         &   1.5E-01         &      0.53         &   $-$         &   Not detected        \\
              \cdsop         &              \vcero         &             \dosuno         &   $\leq$\,1.3E-01         &   4.2E-02         &   $\leq$\,4.1E-01         &   1.4E-01         &      0.52         &   $-$         &   Not detected        \\
                 SiO         &              \vcero         &            \unocero         &   $\leq$\,1.2E-02         &   3.9E-03         &   $\leq$\,3.8E-02         &   1.3E-02         &      0.53         &   $-$         &   Not detected        \\
                             &                             &             \dosuno         &   $\leq$\,5.4E-02         &   1.8E-02         &   $\leq$\,1.7E-01         &   5.8E-02         &      0.67         &   $-$         &   Not detected        \\
                             &                             &        \cincocuatro         &   $\leq$\,1.4E-01         &   4.7E-02         &   $\leq$\,$\leq$\,4.6E-01         &   1.5E-01         &      0.54         &   $-$         &   Not detected        \\
\hline
\end{tabular*}
% \\[1ex]

\end{center}
\small
%\vspace{-1mm}
%\textbf{Notes.} Same as in \tab\ref{lineas_acher} but for IRAS\,20056+1834.

\label{lineas_iras20056}
\end{table*}

\begin{table*}[h]
\caption{Radio molecular line survey of \rsp.}
\small
\tiny
%\centering
\vspace{-5mm}
\begin{center}

\begin{tabular*}{\textwidth}{@{\extracolsep{\fill\quad}}ll @{\hspace{0.5ex}} lccccccl}
\hline \hline
\noalign{\smallskip}
%&&&&&\multicolumn{4}{|c|}{}\\
 %\\[-2ex]

 \multirow{2}{*}{Molecule} & \multicolumn{2}{c}{\multirow{2}{*}{Transition}} & I\,(peak) & $\sigma$ & $\int I\,dV$ & $\sigma \left(\int I\,dV \right)$  & Sp.\,Res. & $V_{LSR}$ &  \multirow{2}{*}{Comments} \\
 &  & & [Jy] & [Jy] & [Jy\,km\,s$^{-1}$] & [Jy\,km\,s$^{-1}$] & [km\,s$^{-1}$]  & [km\,s$^{-1}$] & \\
\hline
\\[-2ex]

               \cdoop         &              \vcero         &             \dosuno         &   $\leq$\,4.4E-02         &   1.5E-02         &   $\leq$\,1.3E-01         &   4.2E-02         &      1.07         &   $-$         &   Not detected        \\
              \cdsop         &              \vcero         &             \dosuno         &   $\leq$\,5.7E-02         &   1.9E-02         &   $\leq$\,1.7E-01         &   5.5E-02         &      0.52         &   $-$         &   Not detected        \\
                 SiO         &              \vcero         &            \unocero         &   7.9E-02         &   8.8E-03         &   0.29         &   2.1E-02         &      0.53         &     56.02         &                       \\
                             &                             &             \dosuno         &   3.5E-01         &   6.8E-03         &   1.6E+00         &   2.8E-02         &      0.67         &     57.02         &                       \\
                             &                             &        \cincocuatro         &   1.3E+00         &   5.6E-02         &   5.5E+00         &   1.4E-01         &      0.27         &     57.50         &                       \\
\hline
\end{tabular*}
% \\[1ex]

\end{center}
\small
%\vspace{-1mm}
%\textbf{Notes.} Same as in \tab\ref{lineas_acher} but for \rsp.

\label{lineas_rsct}
\end{table*}

\end{document}